\documentclass[11pt]{article}
\usepackage{amssymb,amsfonts,amsmath,amsthm,url}
\usepackage{multirow,ctable} % for special table
\usepackage{paralist,xspace, setspace}
\usepackage{simplemargins}
\usepackage{epsfig,graphicx,wrapfig}
\theoremstyle{definition}

\newtheorem{theorem}{Theorem}
\newtheorem*{fact*}{Fact}
\newtheorem*{theorem*}{Theorem}

\newtheorem{example}{Example}[section]
\newtheorem*{example*}{Example}
\newtheorem{exercise}{Exercise}[section]

\newtheorem{conjecture}{Conjecture}
\newtheorem*{conjecture*}{Conjecture}
\newenvironment{definition}[1][Definition.]{\begin{trivlist}
\item[\hskip \labelsep {\bfseries #1}]}{\end{trivlist}}

\def\od{\stackrel{\mathrm{def}}{=}}

\def\RR{\mathbb{R}}

\def\det{\operatorname{det}}

\def\sgn{\operatorname{sgn}}
\def\FP{\operatorname{FP}}

\def\vb{{\bf b}}
\def\vx{{\bf x}}

\newcommand\vtheta{{\boldsymbol{\theta}}}

\def\vv{{\bf v}}

\usepackage{multicol}
\usepackage{color}
\definecolor{gold}{rgb}{0.85,.66,0}
\definecolor{cherry}{rgb}{0.9,.1,.2}
\definecolor{burgundy}{rgb}{0.8,.2,.2}
\definecolor{orangered}{rgb}{0.85,.3,0}
\definecolor{orange}{rgb}{0.85,.4,0}
\definecolor{olive}{rgb}{.45,.4,0}
\definecolor{lime}{rgb}{.6,.9,0}
\definecolor{green}{rgb}{.2,.7,0}
\definecolor{darkgreen}{rgb}{.1,.5,0}
\definecolor{grey}{rgb}{.4,.4,.2}
\definecolor{brown}{rgb}{.4,.2,.1}
\definecolor{blue}{rgb}{0,.0, .81}
\definecolor{bluepurple}{rgb}{.3, .0, .7}

\begin{document}

\noindent {\Large \textbf{Predicting neural network dynamics via graphical analysis}}\\
\emph{A book chapter for advanced undergraduates to appear in ``Algebraic and Combinatorial Computational Biology."\ R.\ Robeva, M.\ Macaulay (Eds) 2018.}\\

\vspace{-.15in}
\noindent Katherine Morrison and Carina Curto\\

\vspace{-.15in}

\begin{abstract}
Neural network models in neuroscience allow one to study how the connections between neurons shape the activity of neural circuits in the brain.  In this chapter, we study Combinatorial Threshold-Linear Networks (CTLNs) in order to understand how the pattern of connectivity, as encoded by a directed graph, shapes the emergent nonlinear dynamics of the corresponding network. Important aspects of these dynamics are controlled by the stable and unstable fixed points of the network, and we show how these fixed points can be determined via graph-based rules. We also present an algorithm for predicting sequences of neural activation from the underlying directed graph, and examine the effect of graph symmetries on a network's set of attractors.

\end{abstract}

\tableofcontents

\section{Introduction}
\subsection{Neuroscience background and motivation}
Neurons in the brain have intricate patterns of non-random connections between them.  Indeed, the complexity of connectivity is one of the most important features setting neurons apart from other types of cells in the body.  How does this connectivity shape dynamics?  This question is of particular interest in the study of local recurrent networks, which contain collections of neurons with similar functional properties.  Such networks are found in cortical areas like the mammalian hippocampus and visual cortex, and the role of recurrent -- as opposed to feedforward \cite{AppendixE}--  connectivity serves to shape neural responses into meaningful patterns of activity.  Even in simple models, however, the effects of connectivity on neural activity are poorly understood.

In this chapter, we focus on the Combinatorial Threshold-Linear Network (CTLN) model, first introduced in 2016 \cite{CTLN-paper}.  This is a simplified mathematical model of neural networks that allows us to focus specifically on connectivity as the key ingredient controlling the dynamics.   The emergent dynamics, however, are nonlinear and complex, exhibiting many of the features believed to underlie information processing in the brain.  For example, CTLNs can be multistable, meaning that the network possesses multiple steady states (a.k.a. stable fixed points).  Depending on the initial condition, the activity will evolve to one steady state or another, mimicking decision-making and memory retrieval in the brain.  In this manner, CTLNs are similar to Hopfield networks and other classical attractor neural networks that are popular models for associative memory \cite{Hopfield1, Amit-ANNs}.  Because of their mathematical tractability, however, CTLNs provide a new window into understanding how detailed connectivity influences these processes in the brain. 

CTLNs also exhibit other aspects of nonlinear dynamics that play a functional role in the nervous system.  For example, a network can possess multiple limit cycles or even multiple chaotic attractors.  Limit cycles, in particular, have long been used to model central pattern generators (CPGs) controlling animal locomotion, breathing, or other periodic behaviors \cite{Marder-CPG, CPG-models, ErmentroutTerman}.  The activity of neurons in a limit cycle is often sequential, with neurons taking turns firing in an orderly sequence of activation.  Such sequences have also been observed in higher-level areas, such as the mammalian cortex and hippocampus \cite{Stark-PNAS,Eva-Science,Luczak-PNAS, Buzsaki}.   As an example, consider the problem of remembering a 7-digit phone number, such as 867-5309.  Many people will repeat the number over and over again in their working memory, a process that can be modeled as selecting a limit cycle in a network where neurons representing the various digits fire in a repeating sequence.  How does the connectivity of a network support these kinds of neural functions?  Can one predict the emergent sequences from the structure of the underlying graph?

In this chapter, we will introduce CTLNs and make some of our motivating neuroscience questions more precise.  Next, we will explore how CTLNs can be analyzed as a patchwork of linear systems of ordinary differential equations (ODEs), with the nonlinear behavior emerging from the transitions between adjacent linear regimes.   After that, we will develop a graph-theoretic analysis that enables us to predict various features of the dynamics directly from the underlying connectivity graph.  These results greatly simplify the fixed point analysis from the previous section, and also reveal the remarkable degree to which the combinatorial structure of the graph controls dynamics, irrespective of the model's other parameters.  Finally, we will use these findings to predict sequences from the graph, and study the effect of symmetry on a network's attractors.  The mathematical topics we will visit along the way include concepts from linear algebra, differential equations, dynamical systems, graph theory, and a bit of group theory (in disguise).

%%%%%%%%%%%%%%%%%%%%%%%%%%%
\subsection{The Combinatorial Threshold Linear Network (CTLN) model}
The dynamics of threshold-linear networks (TLNs) are governed by the following system of ODEs,
\begin{equation}\label{eq:network}
\dfrac{dx_i}{dt} = -x_i + \left[\sum_{j=1}^n W_{ij}x_j+\theta \right]_+, \quad i = 1,\ldots,n
\end{equation}
where $n$ is the number of neurons.
The dynamic variable $x_i(t) \in \RR_{\geq 0}$ is the activity level (or ``firing rate'') of the $i^\textrm{th}$ neuron, and $\theta>0$ is a constant external input.  The values $W_{ij}$ are entries of an $n \times n$ matrix of real-valued connection strengths.   
The threshold nonlinearity $[\cdot]_+ \od \max\{0,\cdot\}$ is critical for the model to produce nonlinear dynamics; without it, the system would be linear (see Appendix for brief review of linear systems of ODEs).

CTLNs are a special case of TLNs, where we restrict to connection strengths $W_{ij}$ that are obtained from a simple\footnote{A graph is \emph{simple} if it does not have loops or multiple edges in the same direction between a pair of nodes.} directed graph $G$ in the following way:

\vspace{-.1in}
\begin{equation} \label{eq:binary-synapse}
W_{ij} = \left\{\begin{array}{cc} 0 & \text{ if } i = j, \\ -1 + \varepsilon & \text{ if } i \leftarrow j \text{ in } G,\\ -1 -\delta & \text{ if } i \not\leftarrow j \text{ in } G. \end{array}\right. \quad \quad \quad \quad
\end{equation}
where $ i \leftarrow j$ indicates that there is an edge from $j$ to $i$ in the graph $G$, and $i \not\leftarrow j$ indicates that there is no such edge.  A CTLN is thus completely specified by the choice of a directed graph $G$, along with three positive real parameters: $\varepsilon, \delta,$ and $\theta$.  We additionally require that $\delta >0$, and $0 < \varepsilon < \frac{\delta}{\delta+1}$; when these conditions are met, we say the parameters are within the \emph{legal range}. 

\begin{figure}[!ht]
\begin{center}
\includegraphics[height=3.05in]{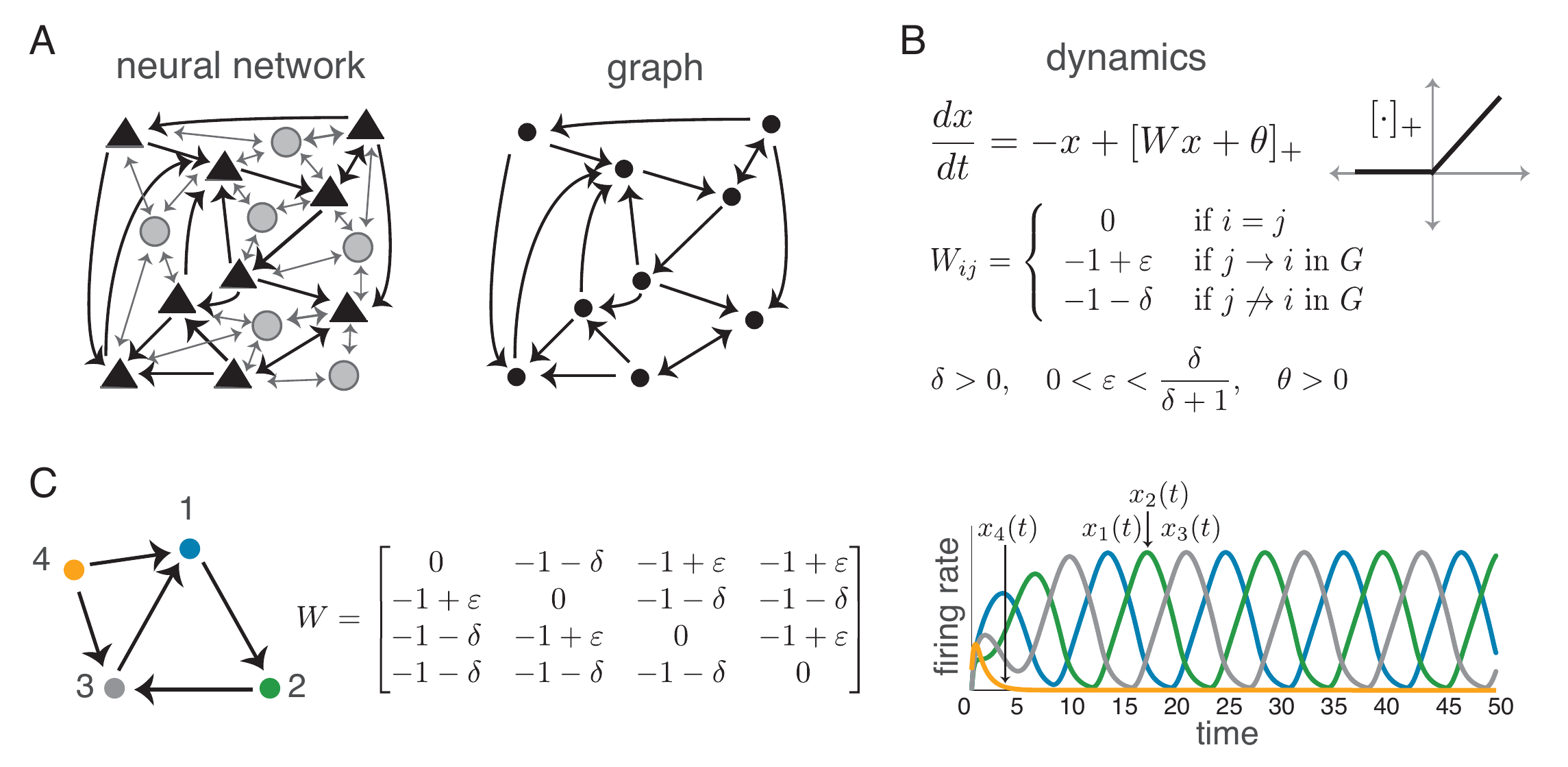}
\vspace{-.2in}
\caption{(A) A neural network with excitatory pyramidal neurons (triangles) and a background network of inhibitory interneurons (gray circles) that produce a global inhibition. The corresponding graph (right) retains only the excitatory neurons and their connections.  (B) Equations for the CTLN model.  (C) A directed graph (left), and its corresponding connection strength matrix $W$ (middle).  (Right) The periodic firing pattern produced by the CTLN with standard parameters.  Firing rate curves are color-coded to match the corresponding neuron in the graph.
}
\label{fig:network-setup-and-3cycle}
\end{center}
\vspace{-.2in}
\end{figure}

The rate of change $dx_i/dt$ consists of two parts: a leak term, $-x_i,$ and a thresholded term.  In the thresholded term, $\sum_{j=1}^n W_{ij}x_j$ is the sum of the weighted synaptic inputs to neuron $i$.  Note that since $x_j\geq 0$, this sum is negative; however, the external drive $\theta>0$ allows the net input to be positive when the inhibitory inputs to the neuron are not too strong.  The leak term ensures that in the absence of a net positive input, when the thresholded term is zero, a neuron's firing rate will decay exponentially to zero.

Notice that $W_{ij} < 0$ whenever $i \neq j$.  We interpret the CTLN as modeling a network of $n$ excitatory neurons, whose net interactions are inhibitory due to a background inhibition that does not enter explicitly into the model (see Figure~\ref{fig:network-setup-and-3cycle}A). When $j \not\to i$, we say $j$ \emph{strongly inhibits} $i$; when $j \to i$, we say $j$ \emph{weakly inhibits} $i$, and we interpret the weak inhibition as the sum of an excitatory connection with the background inhibition.  
Note that because $-1-\delta < -1 < -1+\varepsilon$, when $j \not\to i$, neuron $j$ inhibits $i$ \emph{more} than it ``inhibits itself" via its leak term; when $j \to i$, neuron $j$ inhibits $i$ \emph{less} than it inhibits itself.  These differences in inhibition strength cause the activity to follow the arrows of the graph (see Figure~\ref{fig:network-setup-and-3cycle}C).

For fixed parameters, only the graph $G$ varies between networks.  Thus, we can attribute all differences in dynamics to differences in connectivity, providing insight into how neural connectivity shapes emergent dynamics.  
For all simulations in this chapter, we fix the parameters at $\theta=1$, $\varepsilon=0.25$, and $\delta=0.5$, unless otherwise noted.  We refer to these values as the \emph{standard parameters}.  

\paragraph{Variety of dynamics of CTLNs.} Despite the simplicity of the nonlinearity, CTLNs exhibit the full range of nonlinear dynamic phenomena: multistability, limit cycles, quasiperiodic attractors, and chaos.  Multistability, i.e.\ the coexistence of multiple stable fixed points, is the only nonlinear behavior that occurs in the case where $W$ is symmetric \cite{HahnSeungSlotine, pattern-completion}.  For non-symmetric $W$, limit cycles, chaotic, and quasiperiodic attractors can also occur.  As an example of limit cycle behavior, consider the CTLN in Figure~\ref{fig:network-setup-and-3cycle}C.  Notice that the firing rate curve for the \emph{source} neuron 4 (yellow) quickly decays to 0, and the network settles into a sequential firing pattern following the 3-cycle $(123)$ in the graph.  This limit cycle emerges for every initial condition, and is thus a \emph{global attractor}. 

\begin{figure}[!ht]
\includegraphics[width=\textwidth]{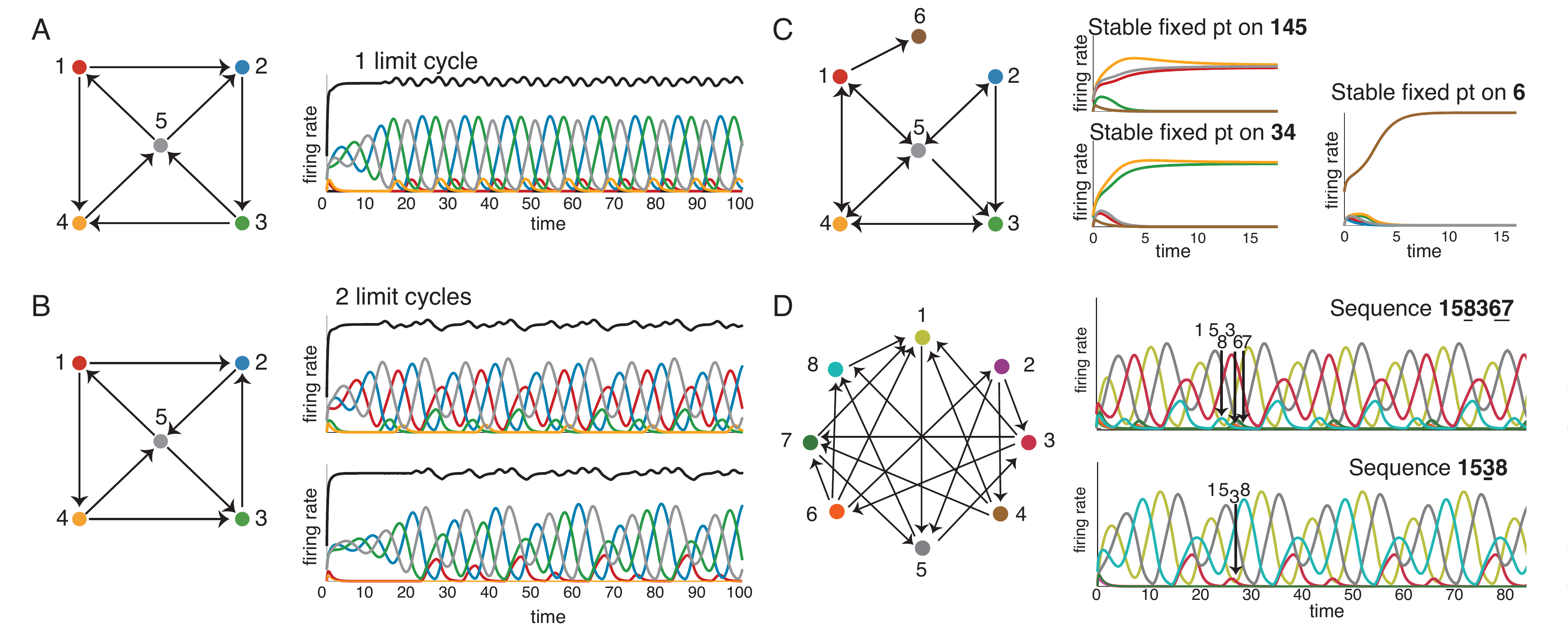}
\vspace{-.2in}
\begin{center}
\caption{Four example graphs with all attractors of the corresponding CTLNs.  Firing rate curves are color-coded to match the corresponding neuron in each graph; the black curves in A and B show total population activity, obtained by summing all the individual firing rates.
}
\label{fig:intro-examples}
\end{center}
\vspace{-.2in}
\end{figure}

Figures~\ref{fig:intro-examples}A and B show two graphs with matching \emph{in-degree} and \emph{out-degree} across the nodes (see box of graph theory terminology in Section~\ref{sec:graph-terms}), and yet they exhibit significantly different dynamics, with A producing a single limit cycle that is a global attractor, while B has two limit cycles.  This shows that the CTLN model exhibits dynamics that are truly \emph{emergent}, as the difference cannot be explained by local properties of the nodes.  Notice that the high-firing neurons in these limit cycles correspond to 3-cycles in the graph.  In graph A, the 3-cycle is $(235)$; in graph B, they are $(125)$ and $(253)$.  Interestingly, both graphs A and B contain an additional 3-cycle $(145)$ that does not have a corresponding limit cycle.  

To any limit cycle, we can associate a sequence of neural firing as shown in Figure~\ref{fig:intro-examples}D.  These sequences are shaped by cycles in the graph, but neurons not involved in a cycle still participate in the sequence, and a given CTLN can produce multiple sequences of different lengths.  Finally, Figure~\ref{fig:intro-examples}C shows that multiple stable fixed points can arise in the same network.  Moreover,
the sets of active neurons, i.e.\ the \emph{supports} of the fixed points, can have different sizes.  Note that each fixed point support corresponds to a \emph{clique} in the graph, but not every clique has a corresponding fixed point.  

This variety of dynamic behaviors motivates a number of questions.  Which graph structures correspond to stable fixed points of the network?  When will limit cycles, chaotic, or quasiperiodic attractors emerge?  Why do some 3-cycles in a graph have corresponding limit cycles, but not others?  What determines the sequence of neural firing in a dynamic attractor?  The primary goal of this chapter is to introduce methods for analyzing the underlying graph in order to predict features of the dynamics.  This will allow us to directly relate a network's connectivity to its dynamics.

%%%%%%%%%%%%%%%%%%%%%%%%%%%
%\FloatBarrier
\section{A CTLN as a patchwork of linear systems}
The dynamics in equation~\eqref{eq:network} can be written more compactly as $\frac{d\vx}{dt} = -\vx +[W\vx +\vtheta]_+$.  If the threshold nonlinearity were dropped, this would yield the linear system $\frac{d\vx}{dt} =(-I+W)\vx +\vtheta$.  Assuming $-I+W$ is invertible, this system has a unique fixed point that is stable if all eigenvalues of $-I+W$ have negative real part, and is unstable otherwise (see Appendix for a brief review of fixed points of linear systems).  
While the threshold is crucial for producing the nonlinear dynamics observed in CTLNs, the fact that the nonlinearity is piecewise linear allows us to analyze these networks as a patchwork of linear systems that partition the positive orthant.  In particular, we can identify and classify the fixed points of a CTLN by analyzing the fixed points of each linear system in the patchwork. 

Let $$y_i \od \sum_{j=1}^n W_{ij}x_j+\theta,$$ and rewrite $\frac{dx_i}{dt} = -x_i +[y_i]_+$.  When $y_i \leq 0$, we obtain $\frac{dx_i}{dt}=-x_i$ for neuron $i$. When $y_i>0$, we have $\frac{dx_i}{dt}=-x_i + y_i$.  Thus the set of hyperplanes $\{y_i=0\}$ partitions the positive orthant into chambers where purely linear systems of ODEs apply.\footnote{The hyperplanes $y_i=0$ should not be confused with the \emph{nullclines} $\frac{dx_i}{dt}=0$.}  We identify each chamber by a corresponding subset $\sigma \od \{i~|~y_i>0\}$; there are up to $2^n$ possible chambers in the positive orthant.  The linear system for each chamber has a fixed point $\vx^*$; this fixed point has the form $x_k^*= 0$ for all $k \notin \sigma$ and $\vx_\sigma^*=(I-W_\sigma)^{-1}\vtheta_\sigma$, where the subscript $\sigma$ indicates restricting the vector/matrix to only the entries indexed by $\sigma$.   Note that $\vx^*$ may or may not be located inside the chamber in which the linear system applies.  Thus, for each $\sigma$, the corresponding fixed point $\vx^*$ is only a true fixed point of the CTLN if it resides in the appropriate chamber. For example, the fixed point $\vx^*=\bf{0}$, corresponding to $\sigma = \emptyset$, will never lie in its corresponding chamber because $\theta >0$, and thus is never a fixed point of a CTLN.  Note a fixed point of a CTLN is stable precisely when all eigenvalues of $-I+W_\sigma$ have negative real part. 

At a fixed point of the CTLN we must have $x_i=[y_i]_+$ for each $i \in [n]$, so for a fixed point $\vx^*$ of the linear system associated to $\sigma$ to be in its correct chamber, the system must satisfy
(i) $x_i^*>0$ for all $i \in \sigma$, 
and (ii)  $y_k^* \leq 0$ for all $k \notin \sigma$, where $y_k^*$ is obtained by evaluating $y_k$ at $\vx^*$.  We refer to (i) and (ii) as the ``on''- and ``off''-neuron conditions, respectively.  When these conditions are all satisfied, we say the CTLN has a fixed point with \emph{support} $\sigma$.

\begin{figure}[!ht]
\begin{center}
\includegraphics[height=2.75in]{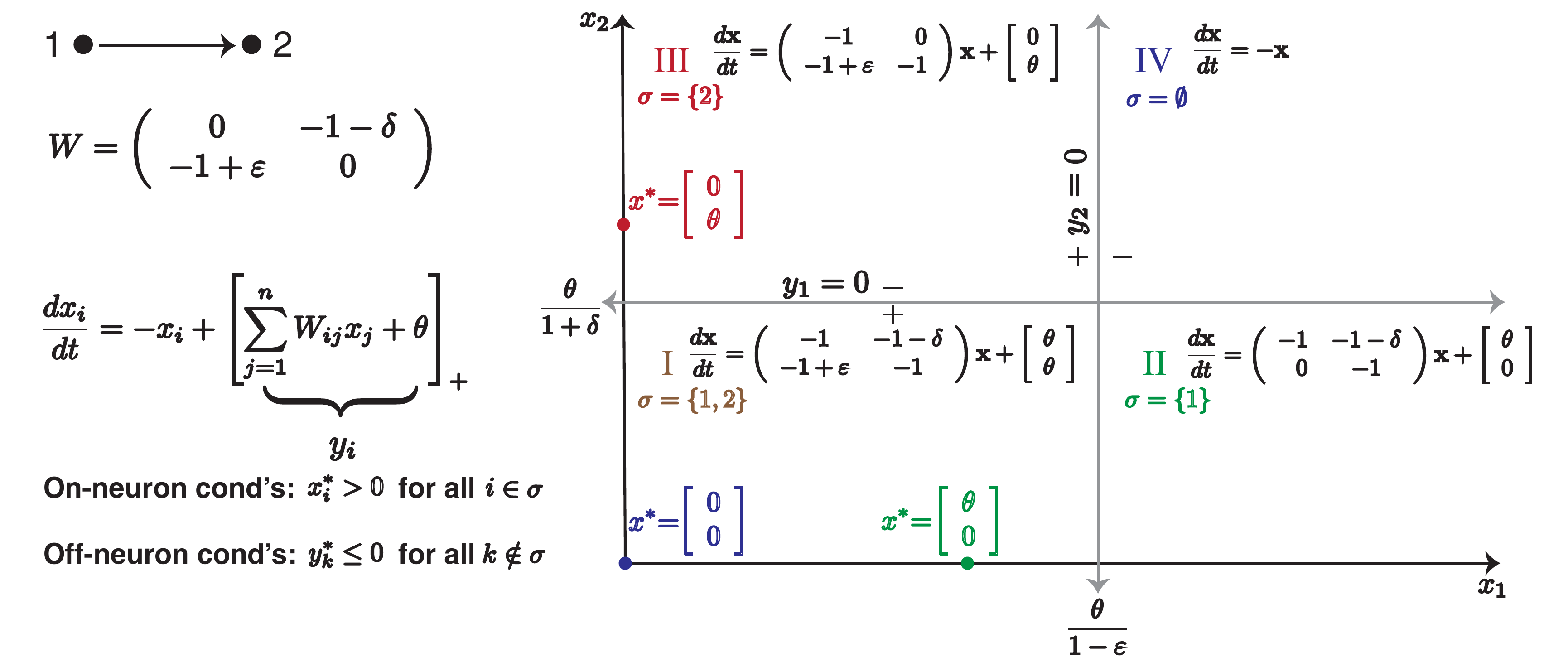}
\vspace{-.1in}
\caption{Patchwork of linear systems for the CTLN corresponding to the graph shown.}
\vspace{-.2in}
\label{fig:single-edge}
\end{center}
\end{figure}

\begin{example}\label{ex:single-edge}
Consider the graph $G$ on two neurons with a single directed edge $1 \to 2$, and the corresponding CTLN (see
 Figure~\ref{fig:single-edge}).
For this network, $y_1=(-1-\delta)x_2 + \theta$ and $y_2=(-1+\varepsilon)x_1 + \theta$.  Thus the hyperplane $y_1=0$ is the horizontal line $x_2=\frac{\theta}{1+\delta}$, and the hyperplane $y_2=0$ is the vertical line $x_1=\frac{\theta}{1-\varepsilon}$.  This cuts the first quadrant into four chambers with a different linear system of ODEs holding in each one.  Each chamber has an associated $\sigma = \{i~|~y_i>0\}$.  

To find the fixed points of the CTLN, we solve for the fixed point $\vx^*$ of each linear system and determine whether it lives in its corresponding chamber.  For chamber I, since $y_1,y_2>0$, we obtain $\frac{dx_i}{dt}=-x_i + W_{ij}x_j+\theta$ for $i=1,2$.  Solving $\frac{d\vx}{dt}=0$ yields the fixed point $$\vx^*=(I-W)^{-1}\left[\begin{array}{r} \theta\\ \theta \end{array}\right] = \dfrac{1}{\delta-\varepsilon(\delta+1)}\left[\begin{array}{r} \delta\theta\\ -\varepsilon\theta \end{array}\right].$$  
Within the legal parameter regime, $x_1^*>0$ and $x^*_2<0$, and so this fixed point violates the on-neuron conditions and lies outside of chamber I.  We conclude that the fixed point for the linear system in chamber I is {\em not} a fixed point of the CTLN.

In contrast, for chamber III, the fixed point $\vx^*=[0,~\theta]^\top$ of the linear system {\em does} lie in its chamber (see Figure~\ref{fig:single-edge}), and is thus a fixed point of the CTLN. By analyzing the remaining two linear systems for chambers II and IV, as in the exercise below, we see that $[0,~\theta]^\top$ is in fact the unique fixed point for the CTLN.  Furthermore, the eigenvalues of the associated matrix for chamber III are both $-1$, and so this fixed point  is stable. 

\begin{exercise}
Verify that the linear systems given for chambers II and IV are those shown in Figure~\ref{fig:single-edge}, and then find the fixed points of those systems.  Show that these fixed points do not lie in their corresponding chambers.  Conclude that these are \underline{not} fixed points of the CTLN.
\end{exercise}
\end{example}

\begin{exercise}
Let $G$ be the graph on two neurons with no edges between them.  Analyze the corresponding CTLN to verify that the network has exactly two stable fixed points, $[\theta,~0]^\top$ and $[0,~\theta]^\top$, and one unstable fixed point $\frac{1}{\delta+2}[\theta, \theta]^\top$.
\end{exercise}

\begin{exercise}
Let $G$ be the graph on two neurons with a bidrectional edge between them.  Analyze the corresponding CTLN to verify that the network has exactly one fixed point, $\frac{1}{2-\varepsilon}[\theta,~\theta]^\top$, which is stable.
\end{exercise}

\subsection{How graph structure affects fixed points}

\begin{figure}[!ht]
\vspace{-.2in}
\begin{center}
\includegraphics[width=.85\textwidth]{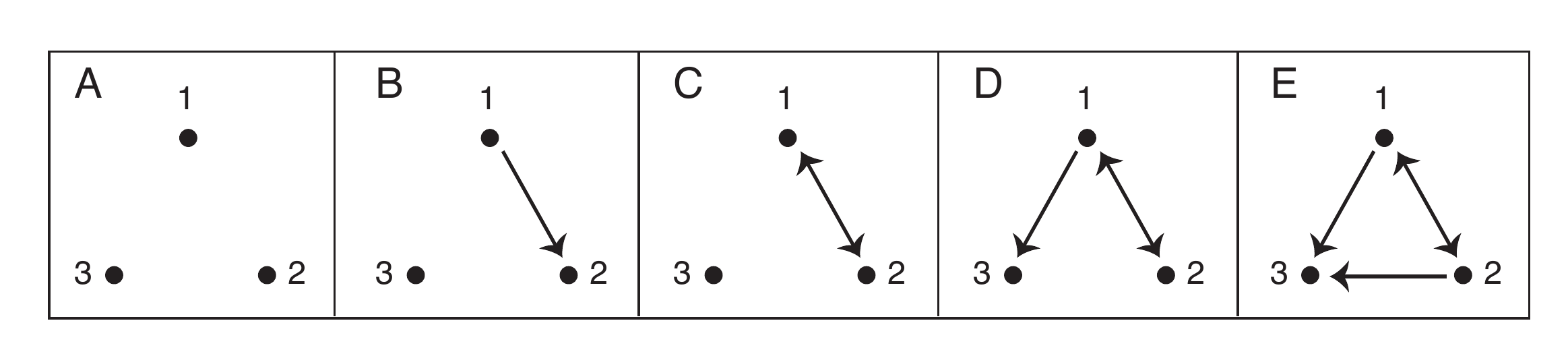}
\vspace{-.2in}
\caption{A sequence of graphs where consecutive graphs differ only by a single edge.}
\label{fig:5-nested-graphs}
\vspace{-.1in}
\end{center}
\end{figure}

To build intuition for how the graph structure affects fixed points, we will compute the fixed points for each of the CTLNs defined by the five graphs in Figure~\ref{fig:5-nested-graphs}, using the same strategy that we used in Example~\ref{ex:single-edge}.  Each graph in this sequence is obtained from the previous one by adding a single edge.  Thus, this analysis will illustrate the impact of individual edges on the collection of CTLN fixed points.  

\begin{figure}[!ht]
\vspace{-.5in}
\begin{center}
\includegraphics[width=.92\textwidth]{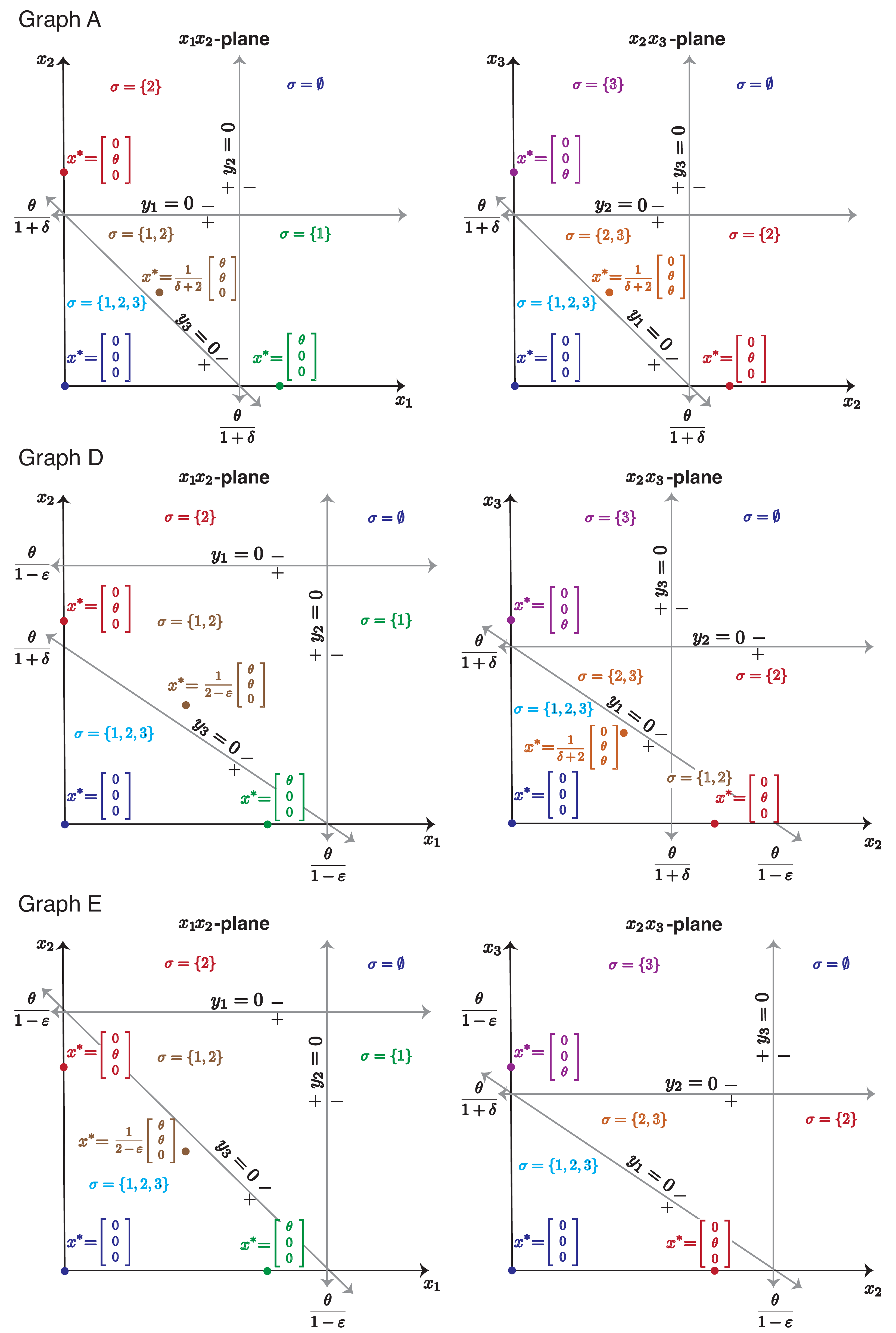}
\vspace{-.2in}
\caption{Projections of chambers of linear systems with their corresponding fixed points for the CTLNs defined by graphs A, D, and E from Figure~\ref{fig:5-nested-graphs}.}
\label{fig:5-phase-planes}
\end{center}
\vspace{-.5in}

\end{figure}

For the CTLN corresponding to graph A, the relevant chambers are defined by the hyperplanes $x_j +x_k =  \frac{\theta}{\delta+1}$ for distinct $j,k \in \{1,2,3\}$. In Figure~\ref{fig:5-phase-planes} we depict slices of these chambers corresponding to the $x_1x_2$-plane and the $x_2x_3$-plane.  
Since $x_3=0$ in the $x_1x_2$-plane, the hyperplane $y_1=0$ projects to the horizontal line $x_2=\frac{\theta}{\delta+1}$, with $y_1>0$ below the line and $y_1<0$ above it.  The hyperplane $y_2=0$ projects to the vertical line $x_1=\frac{\theta}{\delta+1}$, and $y_3=0$ projects to the line $x_2 = -x_1+\frac{\theta}{\delta+1}$.
Each chamber has a corresponding $\sigma=\{i~|~y_i>0\}$ that prescribes for which $i$ we can set $[y_i]_+ = y_i$ in the corresponding linear system, while $[y_k]_+ = 0$ for all $k \notin \sigma$. The analogous picture for the $x_2x_3$-plane is also shown in Figure~\ref{fig:5-phase-planes} (top right).

Table~\ref{table:fp-graphA} contains the value of the fixed point for each linear system in the CTLN for graph A, indexed by $\sigma$.  Note that each fixed point can be obtained by solving its corresponding linear system, as in Example~\ref{ex:single-edge}.  Alternatively, the fixed point can be found as 
$$x_k^*=0 \;\; \text{for all} \;\; k \notin \sigma, \;\; \text{and} \;\; \vx_\sigma^* = (I-W_\sigma)^{-1}\vtheta_\sigma,$$ where the subscript $\sigma$ indicates restricting the vector/matrix to only the entries indexed by $\sigma$.  From the top two panels in Figure~\ref{fig:5-phase-planes}, we see each fixed point, other than $[0,0,0]^\top$, lies in its defining chamber for the chambers shown.  Similarly, the fixed points for $\{1,2,3\}$ and $\{1,3\}$ also lie in their respective chambers (not shown).  Thus, each of these is a fixed point of the CTLN.

 \begin{table}[!ht]
\vspace{.05in}
Graph A
\begin{center}
\begin{footnotesize}
\begin{tabular}{l|c|c|c|c|c|c|c|c}
$\sigma$ & $\{1,2,3\}^*$ & $\{1,2\}^*$ & $\{1,3\}^*$ & $\{2,3\}^*$ & $\{1\}^*$ & $\{2\}^*$ & $\{3\}^*$ &$\emptyset$ \\
\hline \hline
$x^*$ & $\dfrac{1}{2\delta+3}\left[\begin{array}{r} \theta \\ \theta \\ \theta \end{array}\right]$ & $\dfrac{1}{\delta+2}\left[\begin{array}{r} \theta \\ \theta \\ 0\end{array}\right]$ & $\dfrac{1}{\delta+2}\left[\begin{array}{r} \theta \\ 0\\ \theta\end{array}\right]$ & $\dfrac{1}{\delta+2}\left[\begin{array}{r} 0\\ \theta \\ \theta \end{array}\right]$ & $\left[\begin{array}{r} \theta \\ 0 \\ 0 \end{array}\right]$ & $\left[\begin{array}{r} 0\\ \theta \\ 0 \end{array}\right]$ & $\left[\begin{array}{r} 0\\ 0\\ \theta \end{array}\right]$ & $\left[\begin{array}{r} 0 \\ 0 \\ 0 \end{array}\right]$ \\
 \end{tabular}
\caption{Fixed points corresponding to each chamber in the patchwork of linear systems for the CTLN defined by graph A in Figure~\ref{fig:5-nested-graphs}.  * indicates that this is also a fixed point of the CTLN. }

\label{table:fp-graphA}
\vspace{-.15in}
\end{footnotesize}
\end{center}
\end{table}

To check stability of these fixed points, it is sufficient to check the eigenvalues of $-I+W_\sigma,$ since the remaining eigenvalues of the full matrix for the $\sigma$ system are all $-1$.  Computing these eigenvalues, we find that the fixed points corresponding to the singletons $\{1\},\{2\},$ and $\{3\}$ are stable, while the matrices $-I+W_\sigma$ for all other systems have at least one positive eigenvalue, and thus their fixed points are unstable.

\begin{exercise}
Analyze the CTLN for graph B in Figure~\ref{fig:5-nested-graphs}, and verify that the network has exactly two stable fixed points, corresponding to $\{2\}, \{3\}$, and one unstable fixed point, corresponding to $\{2,3\}$.
\end{exercise}

\begin{exercise}\label{ex:graphC}
Analyze the CTLN for graph C in Figure~\ref{fig:5-nested-graphs}, and verify that the network has exactly two stable fixed points, corresponding to $\{1,2\}, \{3\},$ and one unstable fixed point, corresponding to $\{1,2,3\}$.
\end{exercise}

To further investigate the effect of adding edges on the set of CTLN fixed points, we next consider graph D from Figure~\ref{fig:5-nested-graphs}.  This graph differs from graph C only by the addition of the $1\to 3$ edge.  The CTLN for graph C has three fixed points corresponding to the subsets $\{1,2\}, \{3\},$ and $\{1,2,3\}$ (see Exercise~\ref{ex:graphC}).  Similarly, as shown in Table~\ref{table:fp-graphD} and the plots in Figure~\ref{fig:5-phase-planes}, the CTLN for graph D has fixed points corresponding to the same supports.  In fact, the values of the fixed points for $\{1,2\}$ and $\{3\}$ are identical to those from graph C, since the $I-W_\sigma$ matrices are identical; while the value of the fixed point for $\{1,2,3\}$ differs.  Thus, the addition of the $1\to 3$ edge did \underline{not} change the fixed point supports.  One can also check that the stability of each fixed point remains the same.

 \begin{table}[!ht]
\vspace{.05in}
Graph D\\
\begin{center}
\begin{footnotesize}
\vspace{-.15in}
\begin{tabular}{l|c|c|c|c|c}
$\sigma$ & $\{1,2,3\}^*$ & $\{1,2\}^*$ & $\{1,3\}$ & $\{2,3\}$ & $\{3\}^*$  \\
\hline \hline
$x^*$ & $\dfrac{1}{3+\delta-\varepsilon}\left[\begin{array}{r} \theta \\ \theta \\ \theta \end{array}\right]$ & $\dfrac{1}{2-\varepsilon}\left[\begin{array}{r} \theta \\ \theta \\ 0\end{array}\right]$ & $\dfrac{1}{\delta-\varepsilon(\delta+1)}\left[\begin{array}{r} \delta\theta\\ 0\\ -\varepsilon\theta \end{array}\right]$ & $\dfrac{1}{\delta+2}\left[\begin{array}{r} 0\\ \theta \\ \theta \end{array}\right]$ & $\left[\begin{array}{r} 0\\ 0\\ \theta \end{array}\right]$  \\
 \end{tabular}
\caption{Fixed points of the linear systems of the CTLN for graph D in Figure~\ref{fig:5-nested-graphs}.  * indicates fixed points of the CTLN.  Fixed points for $\sigma = \{1\},\{2\}, \emptyset$ (not shown) are not fixed points of the CTLN.
}
\label{table:fp-graphD}
\vspace{-.05in}
\end{footnotesize}
\end{center}
\end{table}

Finally, consider graph E, which differs from graph D only by the addition of the $2 \to 3$ edge.  Since the submatrix $I-W_\sigma$ for $\sigma=\{1,2\}$ does not change, the value of the fixed point for this linear system is the same as for graph D (observe that this fixed point is in the same location on the phase plane plots in Figure~\ref{fig:5-phase-planes}).  However, for graph E this is not a fixed point of the CTLN, because the $y_3=0$  hyperplane is shifted as a result of the added $2 \to 3$ edge.  Additionally, the added edge changes the value of the fixed point for $\{1,2,3\}$ such that it is no longer a fixed point support for the new graph E network.  In fact, $\{3\}$ is the only fixed point support of the CTLN for graph E.   

 \begin{table}[!ht]
\vspace{.05in}
Graph E\\
\begin{center}
\begin{footnotesize}
\vspace{-.15in}
\begin{tabular}{l|c|c|c|c|c}
$\sigma$ & $\{1,2,3\}$ & $\{1,2\}$ & $\{1,3\}$ & $\{2,3\}$ &$\{3\}^*$ \\
\hline \hline
$x^*$ & $\dfrac{1}{2\delta-2\delta\varepsilon-\varepsilon}\left[\begin{array}{r} \delta\theta \\ \delta\theta \\ -\varepsilon\theta \end{array}\right]$ & $\dfrac{1}{2-\varepsilon}\left[\begin{array}{r} \theta \\ \theta \\ 0\end{array}\right]$ & $\dfrac{1}{\delta-\varepsilon(\delta+1)}\left[\begin{array}{r} \delta\theta\\ 0\\ -\varepsilon\theta \end{array}\right]$ & $\dfrac{1}{\delta-\varepsilon(\delta+1)}\left[\begin{array}{r} 0\\ \delta\theta\\ -\varepsilon\theta \end{array}\right]$ & $\left[\begin{array}{r} 0\\ 0\\ \theta \end{array}\right]$   \\
 \end{tabular}
\caption{Fixed points of the linear systems of the CTLN for graph E in Figure~\ref{fig:5-nested-graphs}.  * indicates fixed points of the CTLN.  Fixed points for $\sigma = \{1\},\{2\}, \emptyset$ (not shown) are not fixed points of the CTLN.
}
\label{table:fp-graphE}
\vspace{-.15in}
\end{footnotesize}
\end{center}
\end{table}

From the above analyses, we see that the addition of the $1\to 3$ edge from graph C to graph D does not affect the fixed point supports.  In contrast, the addition of the $2 \to 3$ edge from graph D to graph E dramatically alters the set of CTLN fixed point supports.  Another important point is that the fixed point supports in each case are independent of the values of $\varepsilon$ and $\delta$, provided these fall within the legal range.

Which subgraphs correspond to fixed point supports?  How does the way a subgraph is embedded in the larger graph determine whether or not a fixed point for a linear system survives to be a fixed point of the CTLN?  In the next section, we will focus on the development of graph rules for predicting fixed point supports, thus eliminating the need to do an (often tedious) analysis of the patchwork of linear systems, as we have done here.

%%%%%%%%%%%%%%%%%%%%%%%%%%%
\vspace{-.1in}
\section{Graphical analysis of stable and unstable fixed points}
\vspace{-.05in}

The previous section illustrated how the graph structure controls the collection of CTLN fixed point supports.  Furthermore, the values of $\varepsilon$ and $\delta$ did not affect the set of supports, only the values of the fixed points themselves. But these values can be immediately computed once the collection of fixed point supports is known, since for each support $\sigma$ the corresponding fixed point is given by $\vx^*_\sigma=(I-W_\sigma)^{-1}\vtheta_\sigma$ and $x_k^*=0$ for all $k \notin \sigma$.  We can thus restrict our attention to finding the collection of (stable and unstable) fixed point supports given a graph $G$, which we denote:

\vspace{-.23in}
$$\FP(G) = \FP(G, \varepsilon, \delta) \od \{ \sigma \subseteq [n] ~|~ \sigma \text{ is the support of a fixed point of the CTLN } W(G, \varepsilon, \delta)\}.$$
\vspace{-.23in}

\noindent We will use the notation $\FP(G)$ because all of the following results on fixed point supports are independent of the actual values of $\varepsilon$ and $\delta$, provided these parameters lie within the legal range.

The rest of this section is dedicated to developing tools for finding $\FP(G)$ through graphical analysis alone, without appealing to computations such as the ones performed in Section 2.  All of the mathematical results presented here are contained in~\cite{fp-paper}.

%%%%%%%%%%%%%%%%%%%
\vspace{-.03in}
\subsection{Graph theory concepts}\label{sec:graph-terms}
To aid in the graphical analysis of fixed point supports, the following table reviews background graph theory terminology and some useful new graph concepts. 
\vspace{.15in}
 
 \fbox{
\begin{minipage}{41em}
\noindent {\bf Graph theory terminology.} Let $G$ be a simple directed graph on $n$ nodes.
\begin{itemize}
\item The \emph{in-degree} of a node is the number of incoming edges it receives.
\item The \emph{out-degree} of a node is the number of outgoing edges it projects.
\item A node is a \emph{sink} if it has out-degree 0.
\item A node is a \emph{source} if it has in-degree 0. A source is called \emph{proper} if it has at least one outgoing edge.
\item A node is \emph{isolated} if it has in-degree 0 and out-degree 0.
\item $G$ is \emph{oriented} if it contains no bidirectional edges.
\item For a subset $\sigma$, the \emph{induced subgraph} $G|_\sigma$ (read as ``$G$ restricted to $\sigma$") is the subgraph consisting solely of nodes in $\sigma$ and the edges between those nodes.
\item A subset $\sigma$ is an \emph{independent set} if there are no edges between any pair of nodes in $\sigma$, i.e.\ if every node is isolated in $G|_\sigma$.
\item A subset $\sigma$ is a \emph{cycle} if there is an ordering of the nodes $1, \ldots, |\sigma|$ such that $G|_\sigma$ has $|\sigma|$ edges, all of the form $i \to i+1 \pmod{|\sigma|}$.  
\item A subset $\sigma$ is a \emph{clique} if every pair of nodes in $\sigma$ has a bidirectional edge between them in $G|_\sigma$.  A clique is called \emph{maximal} if it is not contained in any larger clique.
\item A node $k \notin \sigma$ is called a \emph{target of a clique} $\sigma$ if it receives an edge from every node in $\sigma$.  A clique with no targets is called \emph{target-free}.
\item A graph $G$ has \emph{uniform in-degree} $d$ if every vertex has in-degree $d$.
We say that a subset $\sigma$ has \emph{uniform in-degree} $d$ if $G|_\sigma$ has uniform in-degree $d$.
\end{itemize}
\end{minipage}
}

\begin{example}
Consider the graphs in Figure~\ref{fig:5-nested-graphs}.  In graph B, node 1 is a proper source, node 2 is a sink, and node 3 is isolated (it is also a source and a sink).   Graphs A and B are oriented.  In graph A, $\{1,2,3\}$ and all of its non-empty subsets are independent sets, while in graph B, $\{1,3\}$ and $\{2,3\}$ (and trivially the singletons $\{i\}$) are independent sets.   Furthermore, node 3 of graph D is \underline{not} a target of the maximal clique $\{1,2\}$, whereas in graph E, node 3 \underline{is} a target of $\{1,2\}$. 
\end{example}

\begin{figure}[!ht]
\vspace{-.2in}
\begin{center}
\includegraphics[width=.65\textwidth]{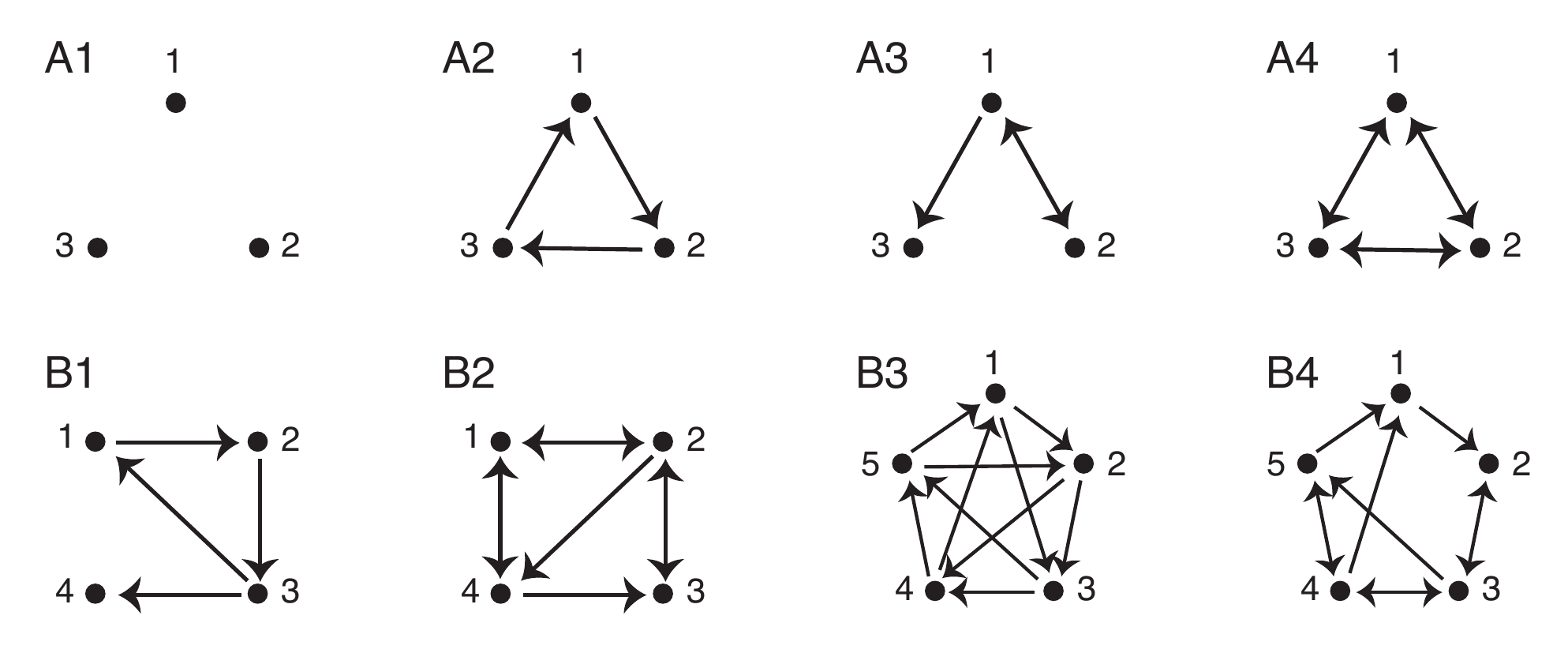}
\vspace{-.1in}
\caption{(A1--4) All uniform in-degree graphs on $n=3$ nodes.  (B1--4) Some other uniform in-degree graphs on $n=4,5$ nodes.}
\label{fig:uniform-in-deg}
\vspace{-.35in}
\end{center}
\end{figure}

\begin{example}
Consider the graphs in Figure~\ref{fig:uniform-in-deg}.  Graph A1 is an independent set and graph A2 is a cycle.  In graph A4, for $\sigma = \{1,2\}$ the induced subgraph $G|_\sigma$ is the pair of nodes 1 and 2 with the double edge between them.  In A4, all subsets of the vertex set $\{1,2,3\}$ are cliques; but only $\{1,2,3\}$ is a maximal clique.  
\end{example}

\begin{exercise}\emph{Maximal vs.\ target-free cliques.}
\vspace{-.07in}
\begin{enumerate}
\item[a.] Prove that all target-free cliques are maximal.  \\

\vspace{-.25in}
\item[b.] Draw an example graph with a maximal clique that is \underline{not} target-free.
\end{enumerate}
\vspace{-.05in}
Conclude that the number of target-free cliques in a directed graph $G$ is less than or equal to the number of maximal cliques in $G$.  
\end{exercise}

Moon and Moser \cite{moon-moser} proved that in an \underline{undirected} graph on $n$ vertices, an upper bound on the number of maximal cliques is given by 
$$ \textrm{max } \# \textrm{ of maximal cliques} = \left\{\begin{array}{cl}3^{n/3} & \textrm{if } n \equiv 0 \pmod 3 \\ 
4\cdot 3^{\lfloor n/3 \rfloor -1} &  \textrm{if } n \equiv 1 \pmod 3 \\
2\cdot 3^{\lfloor n/3 \rfloor} &  \textrm{if } n \equiv 2 \pmod 3 .
\end{array}\right.$$

Given an undirected graph, one can create a corresponding directed graph by replacing each edge with a bidirectional edge.  In this case, every maximal clique of the undirected graph becomes a target-free clique of the directed graph.  Thus, the upper bound on the number of target-free cliques in a directed graph is at least as large as the the upper bound on the number of maximal cliques in an undirected graph.  

\begin{exercise}
For any directed graph, show that there is a corresponding undirected graph such that each target-free clique of the directed graph becomes a maximal clique of the undirected graph.  Conclude that the above upper bound on target-free cliques in a directed graph equals the upper bound on the number of maximal cliques in an undirected graph.
\end{exercise}

\begin{example}
Uniform in-degree subgraphs will prove particularly important in the analysis of fixed point supports.  As examples, if $\sigma$ is a clique, it has uniform in-degree $d=|\sigma|-1$; at the other extreme, if $\sigma$ is an independent set, then it is uniform in-degree with $d=0$.  A cycle has uniform in-degree 1.  Figure~\ref{fig:uniform-in-deg} shows examples of other types of graphs with varying uniform in-degree, including all four uniform in-degree graphs on 3 nodes.  Notice that in a uniform in-degree graph, it is \underline{not} necessary that the graph be symmetric or even that every node have the same out-degree.  Additionally, a subgraph $G|_\sigma$ can have uniform in-degree without every node in $\sigma$ having the same in-degree in $G$; it is only necessary that the nodes have identical in-degrees in the induced subgraph.
\end{example}

\vspace{-.15in}
\begin{exercise}
Draw all graphs on 4 nodes that have uniform in-degree. \emph{Hint:} There are 14 graphs.
\end{exercise}

\vspace{-.15in}
\begin{exercise}
Identify all the uniform in-degree induced subgraphs of graph B4 in Figure~\ref{fig:uniform-in-deg}.  Which of the cliques have targets?  Which are target-free?
\end{exercise}

\vspace{-.2in}
%%%%%%%%%%%%%%%%
\subsection{Stable fixed points}\label{sec:stable-fp}
\vspace{-.1in}

The stable fixed points of a network correspond to steady states, or stable equilibria, of the system.  The population activity $\vx(t)$ can converge to any of the stable fixed points, but which (if any) is selected depends on the initial conditions.  As is typical in attractor neural networks, these fixed points represent stored memory patterns; the process of evolving from an initial condition into one of the stable fixed points is a standard model for \emph{pattern completion} \cite{pattern-completion}.  The set of all stable fixed points is the collection of all static memories stored in the network. 

In this section, we will learn how to infer stable fixed points directly from the graph of a CTLN.  As noted before, we restrict our attention to fixed point supports; recovering the actual values of $\vx^*(t)$ is straightforward once the supports are known. 

Recall that Figure~\ref{fig:intro-examples}C showed how a variety of subgraphs of different sizes could support stable fixed points.  Furthermore, analyzing the networks in Figure~\ref{fig:5-nested-graphs} showed that the interaction of a given subgraph with other nodes in the network affects whether that subgraph corresponds to a fixed point support; for example, the $\{1,2\}$ clique in Figure~\ref{fig:5-nested-graphs}D supports a stable fixed point, while that same clique in Figure~\ref{fig:5-nested-graphs}E does not.  

\vspace{-.15in}

\begin{figure}[!ht]
\begin{center}
\includegraphics[width=.8\textwidth]{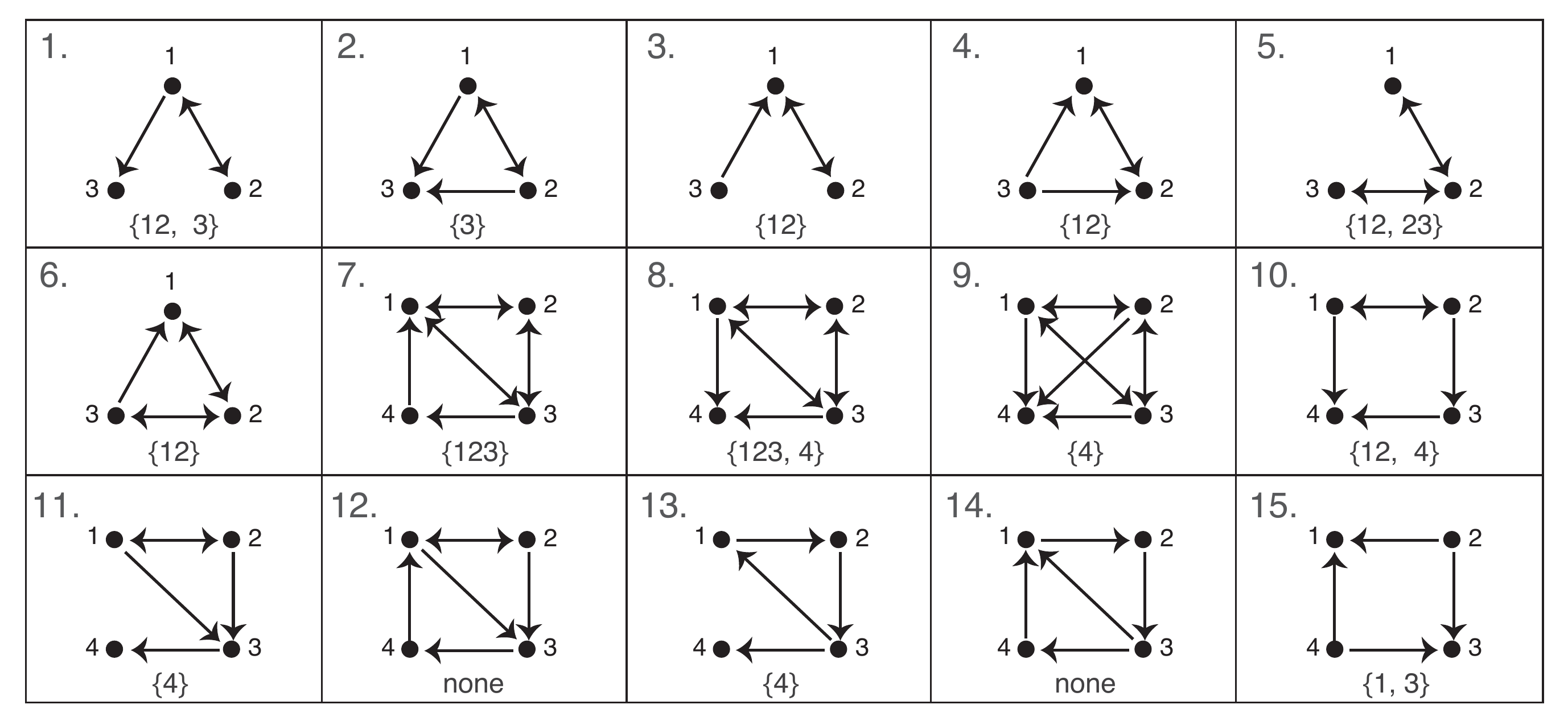}
\vspace{-.15in}
\caption{Graphs for Opening Exploration.  Below each graph is the set of stable fixed point supports for the corresponding CTLN. For example, the stable supports for graph 1 are $\{1,2\}$ and $\{3\}$; we denote the set of these supports as $\{12, 3\}$ for brevity.}
\label{fig:activity-stable-fp}
\end{center}
\vspace{-.2in}
\end{figure}

\paragraph{Opening Exploration: Stable fixed point supports\\}  
\hspace{-.25in} Figure~\ref{fig:activity-stable-fp} shows 15 graphs together with the supports of their stable fixed points.  Carefully analyze these graphs to conjecture which graph structures give rise to stable fixed points.  Be sure to check your conjecture against the full collection of graphs provided.

From the graphs in Figure~\ref{fig:activity-stable-fp}, we see that the only time a single neuron supports a stable fixed point is when it has no outgoing edges, i.e.\ it is a sink.  This fact is actually true more broadly: a single node $i$ is the support of a fixed point for a graph $G$ if and only if $i$ is a sink in $G$; in this case the fixed point is stable.  

When do larger subsets support stable fixed points?  For $\sigma$ to be the support of a stable fixed point, it must be a maximal clique.  However, not every maximal clique is the support of a stable fixed point, as seen in graphs 2, 6, 9, 11 and 12 of Figure~\ref{fig:activity-stable-fp}.  In each of these cases, the cliques that did not support fixed points have a target.  

\begin{fact*}[\cite{fp-paper}]
 A clique $\sigma$ is the support of a fixed point if and only if it is target-free.  In this case, the fixed point is stable.
\end{fact*}

\vspace{-.05in}
Note that a singleton is trivially a clique and is target-free precisely when it is a sink.  Thus the earlier result on sinks is actually a special case of the target-free cliques result.  These are the only subgraphs that have been observed to correspond to stable fixed points, and so we have the following conjecture.

\begin{conjecture}[\cite{CTLN-paper}]
A subset $\sigma \subseteq [n]$ is the support of a \underline{stable} fixed point if and only if $\sigma$ is a target-free clique.
\end{conjecture}

\begin{exercise}
For each of the following sets of possible stable fixed point supports, create a graph whose CTLN has these supports, or explain why no such graph can exist.

\noindent Stable fixed point supports:\\

\vspace{-.1in}
\begin{tabular}{lll}
a) $\{1, 2, 3\}, \ \{3, 4\}$ & b) $\{1, 2, 3\}, \ \{1, 2, 4\}$ & c) $\{1, 2, 3\}, \ \{1, 2, 4\}, \ \{3, 4\}$\\
\\
d) $\{1, 2, 3, 4\} \ \{1, 5\}$ & e) $\{1, 2, 3, 4\} \ \{1, 5\}, \ \{4, 5\}$ & f) $\{1, 2\}, \ \{1, 3\}, \ \{1, 4\}, \ \{2, 3\}$
\end{tabular}
\vspace{.1in}
\end{exercise}

As a special case of the conjecture, we can prove that no stable fixed points emerge for certain classes of graphs guaranteed to have no target-free cliques.

\begin{theorem}[\cite{CTLN-paper, fp-paper}]\label{thm:oriented}
 If $G$ is an oriented graph with no sinks, then the corresponding CTLN has no stable fixed points.  Furthermore, the network dynamics are bounded, and thus are guaranteed to be oscillatory or chaotic.  
\end{theorem}

%%%%%%%%%%%%%%%%%
\subsection{Unstable fixed points}\label{sec:unstable-fp}
Thus far we have focused on stable fixed points, as these produce attractors that may have computational functions such as associative memory storage and retrieval.  But Theorem~\ref{thm:oriented} ensures that if $G$ is an oriented graph with no sinks, then the CTLN has no stable fixed point attractors and can thus exhibit only dynamic attractors: limit cycles, quasiperiodic, or chaotic attractors.  What shapes these attractors?

In Figure~\ref{fig:intro-examples}, the graphs in A and B are oriented with no sinks, and the attractors displayed by these networks are limit cycles.  For the network in Figure~\ref{fig:intro-examples}A, there is a single attractor whose high-firing neurons correspond to the 3-cycle $(235)$.  The graph contains a second 3-cycle $(145)$, yet there is no corresponding attractor.  Meanwhile the network in B, which differs from that in A only by the orientation of the (235) cycle, has attractors corresponding to the 3-cycles $(125)$ and $(253)$, but does not have an attractor corresponding to the 3-cycle $(145)$.  What distinguishes the $(145)$ cycle from the other 3-cycles?  It turns out this can be explained by which 3-cycles support \emph{unstable fixed points}.  Thus, to predict the presence of dynamic attractors, it is essential that we understand what graph structures produce unstable fixed points.  The rest of this section is focused on developing rules for analyzing graphs to find the full collection of fixed point supports $\FP(G)$.  

Recall that for $\sigma$ to support a fixed point, on-neuron conditions must be satisfied for all $i \in \sigma$ and off-neuron conditions must hold for all $k \in [n] \setminus \sigma$.  The on-neuron conditions are independent of any nodes outside of $\sigma$, and thus $\sigma$ satisfies the on-neuron conditions if and only if $\sigma \in \FP(G|_\sigma)$.  Since the off-neuron condition for a given $k \notin \sigma$ can be checked independently of any other nodes outside $\sigma$, this condition is equivalent to checking $\sigma \in \FP(G|_{\sigma \cup \{k\}})$.  This recasting of the fixed point conditions gives Rule~0a in the summary table of rules given below.

\subsubsection{Parity}
Recall that for each support $\sigma$, the fixed point can be computed as $\vx_\sigma^*=(I-W_\sigma)^{-1}\vtheta_\sigma$.  Since each subset has a unique associated fixed point, we have the immediate upper bound $|\FP(G)| \leq 2^n-1$ simply because a network on $n$ nodes has $2^n -1$ nonempty subsets.  This upper bound is attained when $G$ is an independent set (see Exercise~\ref{ex:independent-set}).

Furthermore, as a straightforward consequence of the Poincar\'{e}-Hopf theorem from differential topology \cite{poincare-hopf}, every CTLN must satisfy the following parity condition:
\begin{eqnarray}\label{eq:det-sum}
\displaystyle \sum_{\sigma \in \FP(G)} \sgn \det(I-W_\sigma) &=& 1.
\end{eqnarray}
In particular, since each term in the sum is either $+1$ or $-1$, there must be an odd number of terms and thus $|\FP(G)|$ is odd (Rule~0b: parity).  This fact is particularly useful for determining if there is a fixed point of full support $[n]$ once all proper subgraphs have been analyzed; specifically, $[n] \in \FP(G)$ precisely when there are an even number of smaller subsets that support fixed points.   Equation~\eqref{eq:det-sum} also yields an upper bound on the number of \underline{stable} fixed points of a CTLN.

\begin{exercise}
Use Equation~\eqref{eq:det-sum} to prove that a CTLN on $n$ neurons has at most $2^{n-1}$ stable fixed points.  \emph{Hint:} Recall that a fixed point $\sigma$ is stable if all the eigenvalues of $-I+W_\sigma$ have negative real part.  How are these related to eigenvalues of $I-W_\sigma$ and to its determinant?
\end{exercise}

\subsubsection{Sinks and sources}
To graphically characterize $\FP(G)$, we begin with some of the simplest graph structures to identify, namely \emph{sinks} and \emph{sources}.  Recall that a singleton $\{i\}$ supports a fixed point if and only if it is a sink; in this case the fixed point is stable.  Additionally, an independent set is the support of a fixed point precisely when it is a union of sinks (see Exercise~\ref{ex:union-sinks}).  Finally, a sink $k$ can be added to an existing fixed point support $\sigma$ to create a larger fixed point, as long as $k$ does not ``kill" $\sigma$.  In other words, if $\sigma \in \FP(G)$, so that the sink $k$ satisfied the ``off" neuron conditions and did not kill $\sigma$, then $\sigma \cup \{k\} \in \FP(G)$ as well.  The converse of this also holds, yielding Rule~\ref{rules:sinks}.

\begin{exercise}\label{ex:independent-set}
Prove that if $G$ is an independent set on $n$ nodes, then it has $2^n -1$ fixed points.
\end{exercise}

While sinks are involved in many types of fixed point supports, it turns out that a proper source is \underline{never} involved in a fixed point support.  This holds even when a node is not a source in the full graph, but acts as a proper source in a restricted subgraph.  Specifically, if there exists an $i \in \sigma$ such that $i$ is a proper source in $G|_\sigma$, then $\sigma \notin \FP(G)$.  In fact, if $i$ is a proper source in $G$, then $\FP(G) = \FP(G \setminus \{i\})$ (see Exercise~\ref{ex:sources-rule}).  

\begin{exercise}\label{ex:unidirectional-edge}
Prove that if $\sigma =\{i,j\}$ and $i \to j$ but $j \not\to i$, then $\sigma \not\in \FP(G)$.
\end{exercise}

\fbox{

\begin{minipage}{38em}
\medskip
{\bf Rules for identifying fixed points from graphs} (adapted from \cite{fp-paper})
\begin{enumerate}
\item[0.] \emph{Fixed point conditions and parity.} \label{rules:fp-conditions}
\vspace{-.05in}
\begin{enumerate}
\item A subset $\sigma \in \FP(G)$ $\Leftrightarrow$ $\sigma \in \FP(G|_\sigma)$ and $\sigma \in \FP(G|_{\sigma \cup \{k\}})$ for every $k \notin\sigma.$
\item The total number of (stable and unstable) fixed points, $|\FP(G)|$, is odd. 
\end{enumerate}

\item \emph{Sinks.} \label{rules:sinks}
\vspace{-.05in}
\begin{enumerate}
\item A singleton $\{i\} \in \FP(G)$ $\Leftrightarrow$ $i$ is a sink in $G$.
\item An independent set $\sigma \in \FP(G)$ $\Leftrightarrow$ $\sigma$ is a union of sinks.
\item If $k$ is a sink in $G$,  then $\sigma \cup \{k\} \in \FP(G)$ $\Leftrightarrow$ $\sigma \in \FP(G)$.  
\end{enumerate}

\item \emph{Sources.} \label{rules:sources}
\vspace{-.05in}
\begin{enumerate}
\item If $i \in \sigma$ is a proper source in $G|_{\sigma}$, then $\sigma \notin \FP(G)$.
\item If $i$ is a proper source in $G$, then $\FP(G) = \FP(G \setminus \{i\})$.
\end{enumerate}

\item \emph{Uniform in-degree.} \label{rules:uniform-in-deg}\\
\vspace{-.15in}

\noindent Suppose $\sigma$ has uniform in-degree. Then $\sigma \in \FP(G)$ $\Leftrightarrow$ $\sigma$ is target-free.\\  
\vspace{-.2in}

If $\sigma$ is a target-free clique, then $\sigma$ supports a stable fixed point.
\item \emph{Domination.}  \label{rules:domination}
\vspace{-.05in}
\begin{enumerate}
\item If there exists $j, k \in \sigma$ such that $k$ dominates $j$ w.r.t $\sigma$,\\ then $\sigma \notin \FP(G|_\sigma)$, and so $\sigma \notin \FP(G)$.
\item If there exists $j \in \sigma$ and $k \notin \sigma$ such that $k$ dominates $j$ w.r.t $\sigma$,\\ then $\sigma \notin \FP(G|_{\sigma \cup \{k\}})$, and so $\sigma \notin \FP(G)$.
\item For $j \notin \sigma$, if there exists $k\in \sigma$ such that $k$ dominates $j$ w.r.t $\sigma$,\\ then $\sigma \in \FP(G|_{\sigma \cup \{j\}})$ $\Leftrightarrow$ $\sigma \in \FP(G|_\sigma)$.
\end{enumerate}

\end{enumerate}
\vspace{.025in}
\end{minipage}
}

\subsubsection{Uniform in-degree subgraphs}
We now turn to analyzing when uniform in-degree subgraphs support fixed points.  Recall that a cycle has uniform in-degree, and Figures~\ref{fig:intro-examples}A and 2B showed that some 3-cycles have a corresponding limit cycle, but the 3-cycle $(145)$ never has a corresponding attractor.  This can be explained by the presence/absence of a corresponding unstable fixed point.  To understand when a uniform in-degree subgraph is a fixed point support, we must generalize the notion of target, first introduced for the special case of cliques in Section~\ref{sec:graph-terms}.
Suppose $\sigma$ has uniform in-degree $d$ and $k \notin \sigma$; we say that $k$ is a \emph{target} of $\sigma$ if $k$ receives at least $d+1$ edges from nodes in $\sigma$, and $\sigma$ is said to be \emph{target-free} if it has no targets in $G$.  

\begin{theorem}[\cite{fp-paper}]\label{thm:uniform-in-deg}
Suppose $\sigma \subseteq [n]$ has uniform in-degree $d$ in $G$.  Then $\sigma \in \FP(G)$ if and only if there is no $k \notin \sigma$ receiving at least $d+1$ incoming edges from $\sigma$; in other words, 

\vspace{-.1in}
$$\sigma \in \FP(G) \Leftrightarrow \sigma \textrm{ is target-free in }G.$$ 
If $d < \frac{|\sigma|}{2}$, then the fixed point is unstable.  If $d=|\sigma|-1$, i.e.\ if $\sigma$ is a clique, then it is stable.
\end{theorem}

\begin{figure}[!ht]
\begin{center}
\includegraphics[width=.875\textwidth]{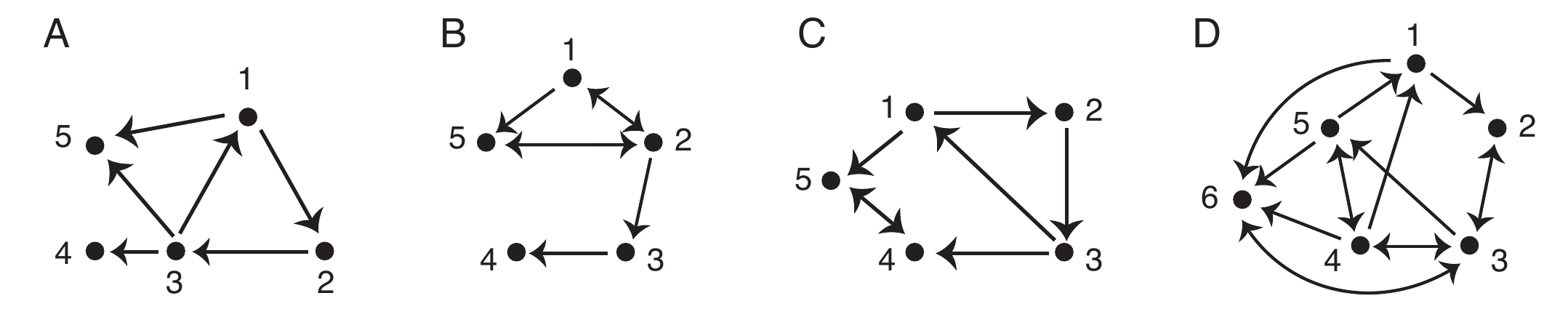}
\vspace{-.2in}
\caption{Uniform in-degree subgraphs with targets; see Example~\ref{ex:target} and Exercise~\ref{exer:target} for details.}
\label{fig:uniform-in-deg-targets}
\vspace{-.15in}
\end{center}
\end{figure}

The intuition behind a target killing a uniform in-degree fixed point with support $\sigma$ is that if all the neurons in $\sigma$ are ``on" at the fixed point, this will force the target node to also turn on since it receives a higher input than any of the nodes in $\sigma$.\footnote{Note that the firing rates of all ``on'' neurons in $\sigma$ are equal at the fixed point if $\sigma$ has uniform in-degree.}  Thus there cannot be a fixed point with only the nodes in $\sigma$ firing.

\begin{example}\label{ex:target} \emph{Targets of uniform in-degree subsets.}\\
Figure~\ref{fig:uniform-in-deg-targets}A contains a 3-cycle $(123)$, which has uniform in-degree $d=1$.  Node 5 is a target of $\{1,2,3\}$ since it receives at least $d+1=2$ edges; in contrast, node 4 is \underline{not} a target since it only receives $d$ edges.  Thus, $\{1,2,3\}$ is not a fixed point support of $G$, although it is a fixed point of $G|_{\{1,2,3,4\}}$ (note also that $\{1,2,3,4\}$ actually has uniform in-degree 1, so it is a fixed point of that restricted subgraph as well).  In Figure~\ref{fig:uniform-in-deg-targets}B, both $\{1,2,3\}$ and $\{1,2,3,4\}$ have uniform in-degree 1, and node 5 is a target of both of these sets, guaranteeing they do \underline{not} support fixed points.  Note that the edge from 5 back to 2 is irrelevant to the uniform in-degree of the induced subgraphs $G|_{\{1,2,3\}}$ and $G|_{\{1,2,3,4\}}$ and is irrelevant to node 5 being a target.  The independent set $\{2,4\}$ has uniform in-degree 0, but since node 2 is not a sink, $\{2,4\}$ does not support a fixed point (see Exercise~\ref{ex:union-sinks}).  
\end{example}

\begin{figure}[!ht]
\begin{center}
\vspace{-.15in}
\includegraphics[width=.5\textwidth]{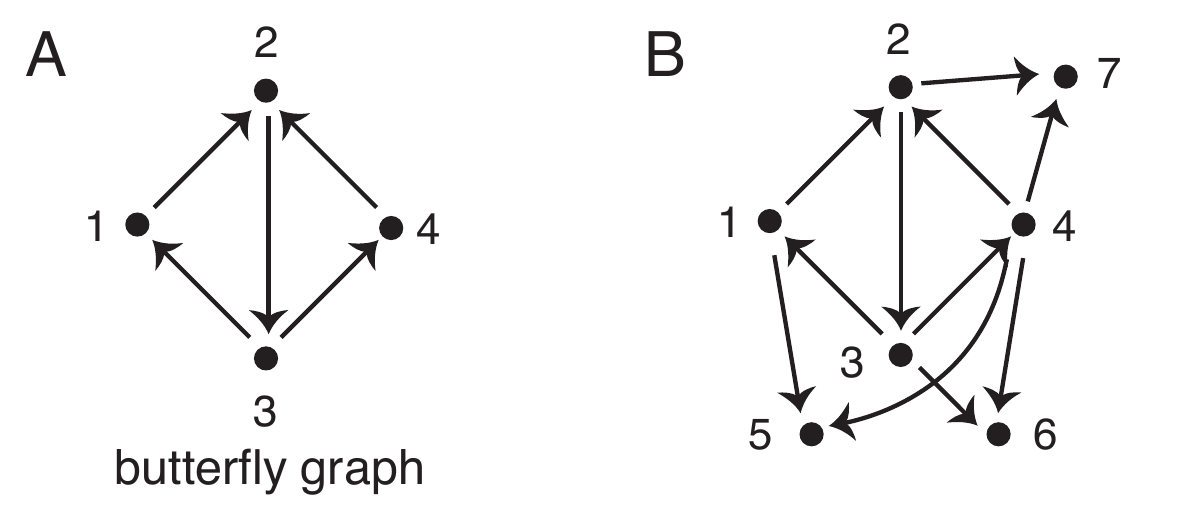}
\vspace{-.15in}
\caption{(A) Butterfly graph.  (B) Example nodes added to the butterfly graph.  }
\label{fig:butterfly-survival}
\vspace{-.3in}
\end{center}
\end{figure}

\begin{example}\label{ex:butterfly-fp}\emph{$\FP(G)$ for the butterfly graph.}\\
Consider the \emph{butterfly graph} shown in Figure~\ref{fig:butterfly-survival}A.  We will identify all the uniform in-degree subgraphs and use this to determine the full set of fixed point supports.  There are two 3-cycles, $(123)$ and $(234)$, which are both uniform in-degree 1.  Neither has an external node receiving two or more edges, so they are both target-free and thus support fixed points.  There is one other uniform in-degree subgraph, the independent set $\{1, 4\}$ with $d=0$.  Node 2 is a target of this set, since it receives at least 1 edge from $\{1, 4\}$; thus $\{1,4\}$ is not a fixed point support.  

Note that no singletons can support a fixed point since there are no sinks (Rule~\ref{rules:sinks}a).  Additionally, no pair of nodes can support a fixed point because every pair other than $\{1,4\}$ has just a unidirectional edge (see Exercise~\ref{ex:unidirectional-edge}).  Every set of three nodes, other than the 3-cycles, contains a node that is a proper source in the subgraph, and so cannot be a fixed point support (Rule~\ref{rules:sources}a).  Thus the only proper subsets that support fixed points are $\{1,2,3\}$ and $\{2,3,4\}$.  But by parity (Rule~0b), $|\FP(G)|$ must be odd, so we conclude that the full support $\{1,2,3,4\}$ must also yield a fixed point.  Thus for the butterfly graph, $\FP(G) = \{123,234,1234\}$; note that for simplicity we drop the set notation for each support contained in $\FP(G)$.
\end{example}

\noindent \begin{exercise}\label{exer:target} \emph{Uniform in-degree and targets.}
\vspace{-.1in}
\begin{enumerate}
\item[a.]  Find all uniform in-degree subsets of the graph in Figure~\ref{fig:uniform-in-deg-targets}C.  Determine which of these are fixed point supports.
\item[b.]  Find all uniform in-degree subsets of the graph in Figure~\ref{fig:uniform-in-deg-targets}D.  Determine which of these are fixed point supports.
\end{enumerate}
\vspace{.05in}
\end{exercise}

\begin{exercise}\label{ex:union-sinks}
\emph{Unions of sinks.}
\vspace{-.1in}
\begin{enumerate}
\item[a.] Using Theorem~\ref{thm:uniform-in-deg}, prove Rule~\ref{rules:sinks}b showing that if $\sigma \subseteq [n]$ is an independent set, then $\sigma \in \FP(G)$ if and only if every node $i \in \sigma$ is a sink in $G$.
\item[b.] Prove that if a graph $G$ has $s$ sinks, then $|\FP(G)| \geq 2^s -1.$
\end{enumerate}
\end{exercise}

\subsubsection{Domination}

The intuition behind Theorem~\ref{thm:uniform-in-deg}, of why a target node 
would be turned on in the presence of a fixed point of uniform in-degree, can be extended to other 
scenarios. This leads us to the concept of \emph{domination}, where a node receives the same inputs as another node, and possibly more.

\begin{definition}
We say that \emph{$k$ dominates $j$ with respect to $\sigma$}, and write $k >_\sigma j$, if $\sigma \cap \{j, k\} \neq \emptyset$ and the following three conditions hold:
\begin{itemize}
\item[(1)] for each $i \in \sigma \setminus \{j, k\}$, if $i \to j$ then $i \to k$,
\item[(2)] if $j \in \sigma$, then $j \to k$, and 
\item[(3)] if $k \in \sigma$, then $k \not\to j$.
\end{itemize}
\end{definition}

\noindent Note that if $k >_\sigma j$, then $j \not >_\sigma k$, and thus $>_\sigma$ is an \emph{antisymmetric} relation.

\begin{exercise} 
Prove that domination $>_\sigma$ is \emph{transitive} for $j,k,\ell \in \sigma$; in other words,  if $\ell >_\sigma k$ and $k >_\sigma j$, then $\ell >_\sigma j$.
\end{exercise}

The following theorem shows how domination can be used to rule in or rule out certain fixed point supports. This gives us Rule~4 in the table of graph rules.

\begin{theorem}[\cite{fp-paper}]\label{thm:domination}
Suppose $k$ dominates $j$ with respect to $\sigma$.  The following statements all hold:
\begin{itemize}
\item[(a)] [inside-in] If $j,k \in \sigma$, then $\sigma \notin \FP(G|_\sigma)$, and so $\sigma \notin \FP(G)$.
\item[(b)] [outside-in] If $j \in \sigma$ and $k \not\in \sigma$, then $\sigma \notin \FP(G|_{\sigma\cup\{k\}})$, and so $\sigma \notin \FP(G)$.
\item[(c)] [inside-out] If $j \notin \sigma$ and $k \in \sigma$, then $\sigma \in  \FP(G|_{\sigma \cup \{j\}})$ if and only if
$\sigma \in \FP(G|_\sigma)$.
\end{itemize}
\end{theorem}

\begin{figure}[!ht]
\vspace{-.3in}
\begin{center}
\includegraphics[width=.75\textwidth]{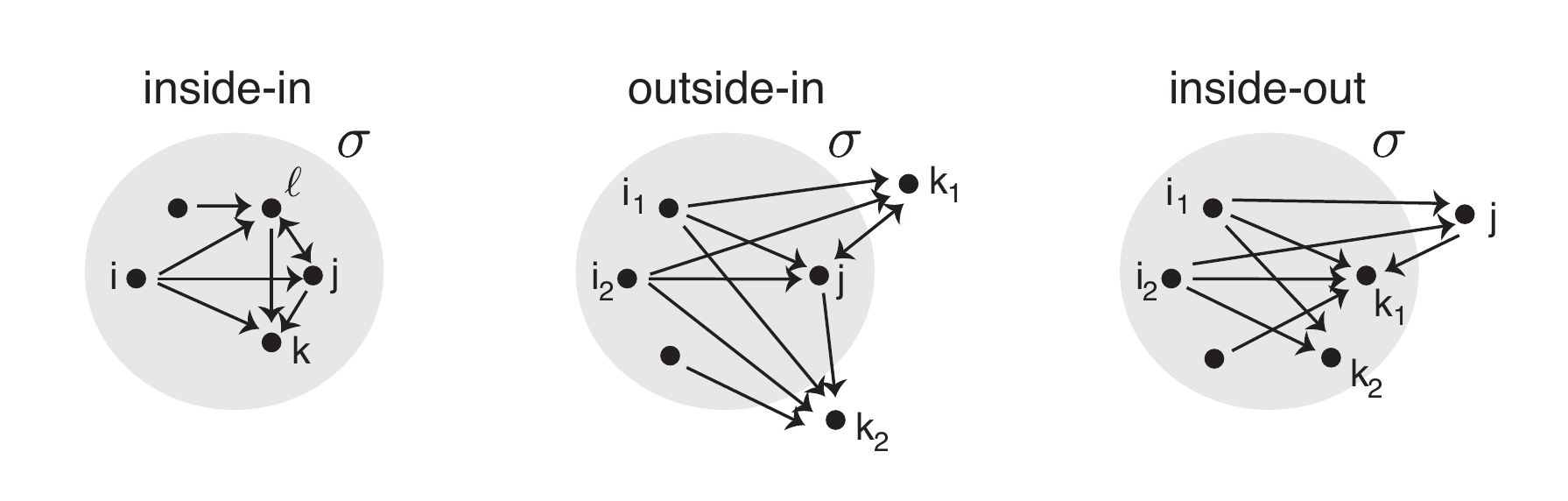}
\vspace{-.3in}
\caption{Three cases of domination.}
\label{fig:domination-examples}
\vspace{-.2in}
\end{center}
\end{figure}

Figure~\ref{fig:domination-examples} illustrates the three cases of domination.  In the first panel, both $k$ and $\ell$ receive all inputs that node $j$ receives (as well as possibly other inputs), and so condition (1) of the definition of domination is satisfied.  Since $j, k, \ell \in \sigma$ we also need $j \to k$ and $k \not\to j$ for domination to hold.  This is true for $k$, and so $k$ dominates $j$; however, this does not hold for $\ell$ since $\ell \to j$, and so $\ell$ does \underline{not} dominate $j$.  A single inside-in domination relationship is sufficient to rule out a fixed point, though, and thus by Theorem~\ref{thm:domination}(a), $\sigma$ is \underline{not} a fixed point support.  The second panel in Figure~\ref{fig:domination-examples} illustrates outside-in domination.  In this case, both $k_1$ and $k_2$ dominate $j$ since both receive all the inputs that $j$ receives from $\sigma$ and both receive an edge from $j$; since $k_1, k_2 \not\in \sigma$ it is not necessary that that $k \not\to j$.  By Theorem~\ref{thm:domination}(b) we conclude that $\sigma$ cannot support a fixed point.  Finally, the third panel shows inside-out domination: $k_1, k_2 \in \sigma$ both receive all inputs that $j$ receives, and there is no edge $k \to j$; thus both $k_1$ and $k_2$ dominate $j$.  Note that it is not necessary that $j \to k$ since $j \notin \sigma$.  Theorem~\ref{thm:domination}(c) guarantees that $\sigma$ survives the addition of node $j$, i.e.\ $\sigma \in \FP(G|_{\sigma \cup \{j\}})$, precisely when $\sigma \in \FP(G|_\sigma)$. 

\begin{example}\label{ex:butterfly-survival}\emph{Butterfly graph survival rules.} 
Returning to the butterfly graph, we examine when the fixed point $\sigma = \{1,2,3,4\}$ survives the addition of a single node.  Note that since $\sigma$ is a fixed point of the butterfly graph, it cannot contain any inside-in domination relationships by Theorem~\ref{thm:domination}(a); thus we need only determine whether inside-out or outside-in domination arise from the addition of the single node.  

In Figure~\ref{fig:butterfly-survival}B, consider $G|_{\{1,2,3,4,5\}}$, where node 5 receives from 1 and 4.  Node 5 does not dominate any nodes in $\sigma$ since 1 and 4 both receive from node 3 while 5 does not.  In fact, node 5 is dominated by node 2, since 2 receives from 1 and 4 and $2 \not\to 5$.  Thus, by Theorem~\ref{thm:domination}(c), inside-out domination guarantees that $\sigma \in \FP(G|_{\{1,2,3,4,5\}})$ since $\sigma$ was a fixed point of the restricted butterfly graph.  

Next, consider $G|_{\{1,2,3,4,6\}}$ where node 6 receives from nodes 3 and 4.  In this case, 6 dominates 4 since 6 receives from 3 and $4 \to 6$.  Theorem~\ref{thm:domination}(b) guarantees that $\sigma \not\in \FP(G|_{\{1,2,3,4,6\}})$ by outside-in domination.  

The survival of $\sigma$ when a single node is added can actually be determined via domination in every case (see Exercise~\ref{exer:butterfly-survival}), except when the added node receives from 1 and 2, or equivalently (by symmetry) from 2 and 4, as in the case of node 7 in Figure~\ref{fig:butterfly-survival}B.  In this case, there are no domination relationships of any type.  Thus an explicit computation is necessary to check whether the added node satisfies the ``off" neuron condition for $\sigma$.  It turns out this condition is in fact satisfied and $\sigma \in \FP(G|_{\{1,2,3,4,7\}})$.  Furthermore, the ``off" neuron condition holds for every value of $\varepsilon$ and $\delta$ in the legal range, and thus the survival rules for the butterfly graph are parameter-independent.  
\end{example}

\begin{exercise}\label{exer:butterfly-survival}
Let $G$ be the butterfly graph union a single node $k$, and let $\sigma =\{1,2,3,4\}$ so that $G|_\sigma$ is the butterfly graph.  
Prove that $\sigma \in \FP(G)$ if and only if $k$ receives at most one edge from $\sigma$ or $k$ receives two edges from $\sigma$ from among the nodes 1, 2, and 4.  \\
\emph{Hint:} Use domination for each of the remaining cases not covered in Example~\ref{ex:butterfly-survival}.
\end{exercise}

\begin{exercise}\label{ex:sources-rule}
Use domination to prove Rule~\ref{rules:sources}:
\begin{enumerate}
\item[(a)] If $i \in \sigma$ is a proper source in $G|_{\sigma}$, then $\sigma \notin \FP(G)$.
\item[(b)] If $i$ is a proper source in $G$, then $\FP(G) = \FP(G \setminus \{i\})$.
\end{enumerate}
\end{exercise}
In addition to not participating in fixed point supports, proper sources appear to ``die" in dynamic attractors as well.  For example, in Figure~\ref{fig:network-setup-and-3cycle}C node 4 is a proper source, and this node is not active once the network falls into its global attractor corresponding to the 3-cycle (123). 
\begin{conjecture}
Proper sources always die and are never active in an attractor of a network.
\end{conjecture}

\begin{exercise}[Mini project]
Consider graph B from Figure~\ref{fig:5-nested-graphs}.  Node 1 is a proper source and thus is not involved in any attractors of the CTLN, so the attractors of graph B are identical to those of the independent set of neurons 2 and 3.  Interestingly, though, the presence of the source 1 affects the basins of attraction of those attractors.  

Using the Matlab code provided with this chapter (see Section~\ref{sec:prediction}), study the CTLN for the independent set on neurons 2 and 3 to confirm that half of the space of initial conditions evolves to the stable fixed point supported on $\{2\}$ while the other half evolves to $\{3\}$, and so the two basins of attraction have the same size.  Next, explore the CTLN for graph B to see how the addition of 1 as a source affects the size of the basins of attraction of $\{2\}$ and $\{3\}$.
\end{exercise}

\begin{exercise}
Suppose $i \in \sigma$ is an isolated node in $G|_\sigma$.  Using domination, prove that if $i$ is \underline{not} a sink in $G$, then $\sigma \notin \FP(G)$.  
\end{exercise}

\begin{exercise}
Prove that an $n$-cycle has a unique fixed point.
\end{exercise}

\begin{exercise}
Explain why if $\sigma$ has uniform in-degree, then there can be no $j, k \in \sigma$ such that $k$ dominates $j$.
\end{exercise}

\subsubsection{Using graph rules to compute $\FP(G)$}
\begin{figure}[!ht]
\vspace{-.2in}
\begin{center}
\includegraphics[height=1.2in]{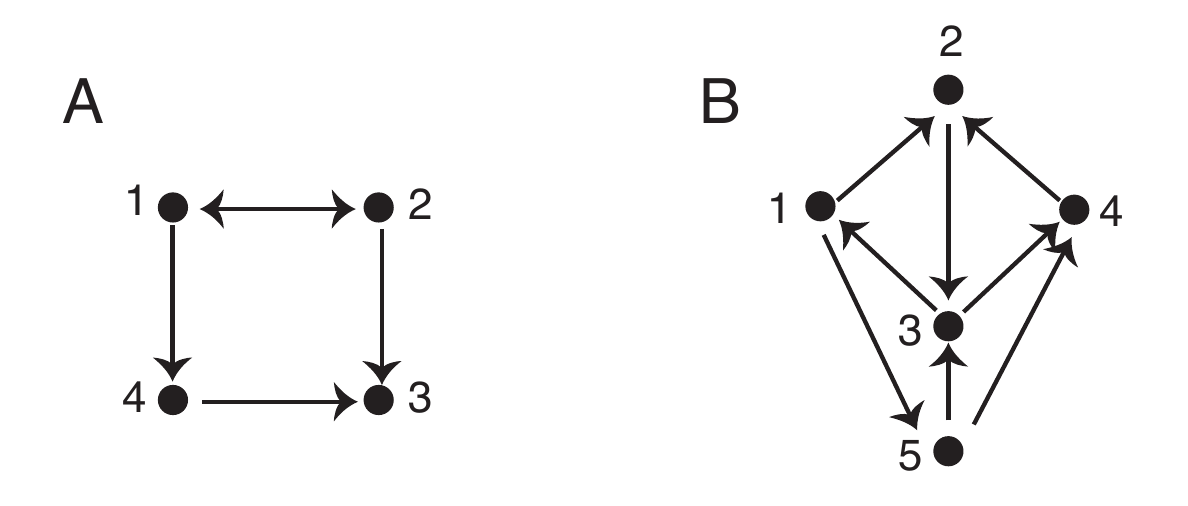}
\vspace{-.2in}
\caption{Graphs for Example~\ref{ex:full-fp}.}
\label{fig:full-fp-examples}
\vspace{-.25in}
\end{center}
\end{figure}

\vspace{-.05in}
As a culmination of the rules in this section, we will demonstrate how to find $\FP(G)$ for the two graphs in Figure~\ref{fig:full-fp-examples}.

\begin{example}\label{ex:full-fp}\emph{Finding $\FP(G)$.}\\
(A) Let $G$ be the graph in Figure~\ref{fig:full-fp-examples}A.  Node 3 is the only sink, and so by Rule~\ref{rules:sinks}(a), $\{3\}$ is the only singleton fixed point support.  The clique $\{1,2\}$ is target-free, and thus supports a fixed point by Rule~\ref{rules:uniform-in-deg}.  The independent sets $\{1,3\}$ and $\{2,4\}$ are not unions of sinks, and thus do not support fixed points by Rule~\ref{rules:sinks}(b).  Every other pair of nodes has a unidirectional edge, yielding a proper source, and so cannot be fixed point supports (Rule~\ref{rules:sources}(a)).  The subset $\{1,2,3\}$ has uniform in-degree 1, and 4 is not a target, hence $\{1,2,3\} \in \FP(G)$ by Rule~\ref{rules:uniform-in-deg}.  Note that $\{1,2,3\}$ is also the union of a fixed point $\{1,2\}$ with a sink, and so it is a fixed point support by Rule~\ref{rules:sinks}(c).  The subset $\{1,2,4\}$ also has uniform in-degree 1, but 3 is a target, and so $\{1,2,4\} \not\in \FP(G)$.  Additionally, $\{2,3,4\} \not\in \FP(G)$ by Rule~\ref{rules:sources}(a) since both 2 and 4 are proper sources in $G|_{\{2,3,4\}}$.  Thus, $\FP(G)$ contains three proper subsets; by Rule~0(b), $|\FP(G)|$ is odd, and so the full set $\{1,2,3,4\} \notin \FP(G)$.  Thus, $\FP(G) = \{3, 12, 123\}$.  \\

\noindent (B)  Let $G$ be the graph in Figure~\ref{fig:full-fp-examples}B.  $G$ has no sinks, so no singletons and no independent sets can be fixed point supports (Rule~\ref{rules:sinks}).  Since $G$ is an oriented graph, every pair of nodes that is not an independent set must have a unidirectional edge, and so contains a proper source and cannot be a fixed point support.  Among the triples, $\{1,2,3\}$ and $\{2,3,4\}$ both have uniform in-degree 1 and are target-free, so they are fixed point supports.  In contrast, $\{1,3,5\}$ has node 4 as a target, and so $\{1,3,5\} \notin \FP(G)$.  All other triples contain a proper source in the restricted subgraph, and thus cannot support fixed points.  Both $\{1,2,3,4\}$ and $\{1,2,3,5\}$ are butterfly graphs, and by the survival rules in Exercise~\ref{exer:butterfly-survival}, $\{1,2,3,4\} \in \FP(G)$ while $\{1,2,3,5\} \notin \FP(G)$.  Both $\{1,2,4,5\}$ and $\{2,3,4,5\}$ contain proper sources in the restricted subgraphs and so do not support fixed points.  The subset $\{1,3,4,5\} \notin \FP(G)$ since 4 dominates 3; additionally, $\{1,3,4,5\}$ is the union of a non-fixed point and a sink, and so cannot support a fixed point by Rule~\ref{rules:sinks}(c).  Thus, $\FP(G)$ contains four proper subsets, and so by parity (Rule~0(b)) the full set $\{1,2,3,4,5\} \in \FP(G)$.  Thus, $\FP(G) = \{123, 234, 1234, 1235, 12345\}$.  
\end{example}

\begin{exercise}
Return to the graphs in the Opening Exploration of Section~\ref{sec:stable-fp} (see Figure~\ref{fig:activity-stable-fp}).  Use the graph rules summarized in the table to find $\FP(G)$ for each graph.  
\end{exercise}

%%%%%%%%%%%%%%%%%%%
\section{Predicting dynamic attractors via graph structure} \label{sec:prediction}
In addition to static memory patterns, which are given by stable fixed points, neural networks also encode dynamic patterns of neural activity.  These are typically modeled by periodic attractors, such as limit cycles, that represent repeating patterns of neural activation.  Such patterns often take the form of sequences, in which neurons fire in a repeatable order that is functionally meaningful.  Such sequences have been observed in the mammalian cortex, hippocampus, and central pattern generator circuits \cite{Marder-CPG, CPG-models,Stark-PNAS,Eva-Science,Luczak-PNAS}.  They model everything from episodic memories (i.e., sequences of places or events) to rhythmic locomotion \cite{ErmentroutTerman, Buzsaki}.  The problem of remembering a phone number, first described in the introduction, is an example of a sequence being maintained in working memory.

Recall from Theorem~\ref{thm:oriented} that if $G$ is an oriented graph with no sinks, then the attractors of the corresponding CTLN will be oscillatory or chaotic.  Typically, given such a dynamic attractor, it is possible to associate a sequence of neural firing based on the order in which neurons achieve their peak firing rate.  
\begin{figure}[!ht]
\begin{center}
\includegraphics[height=1.5in]{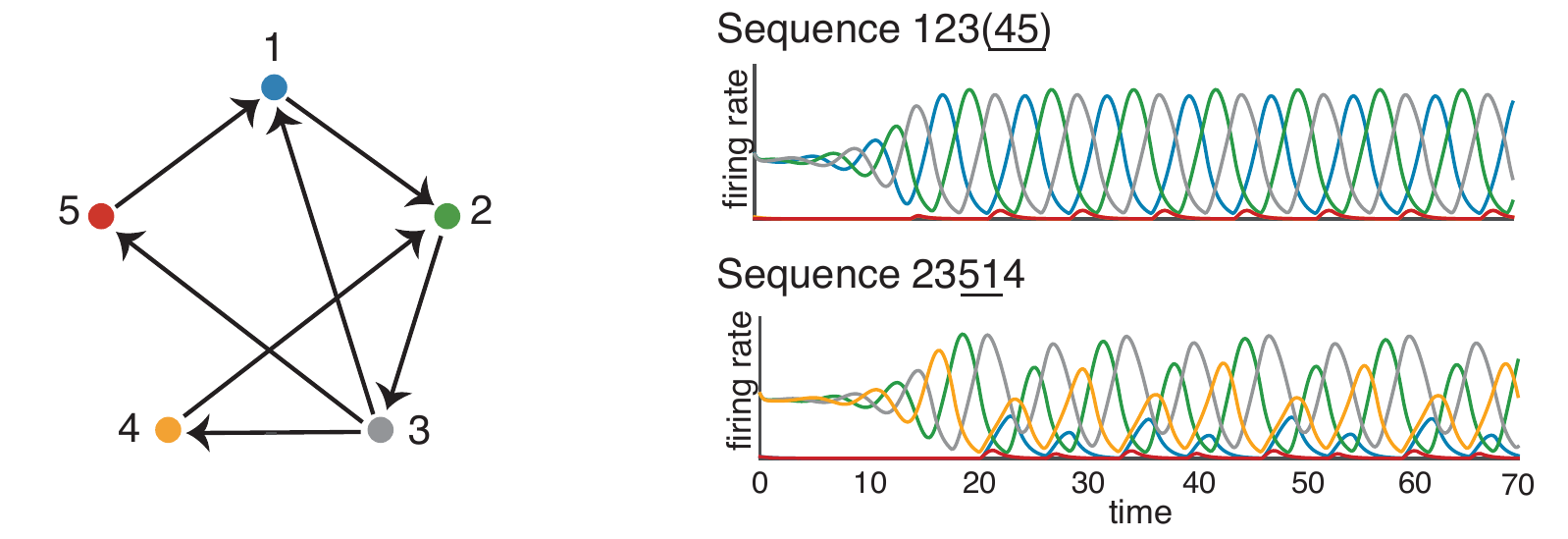}
\vspace{-.2in}
\caption{Graph for a CTLN with two limit cycles, and the corresponding sequences.  In the top limit cycle, neurons 4 and 5 fire synchronously, and so only the firing rate curve for neuron 5 (red) is visible. These two limit cycles are the attractors for parameters $\varepsilon = 0.35$ and $\delta = 0.9$. The top limit cycle has a much larger basin of attraction, while the bottom limit cycle occurs for a smaller set of initial conditions. It can be obtained by initializing the CTLN at or near the unstable fixed point corresponding to $\sigma = \{2,3,4\}$.
%\katie{Note this is graph F3[1]}
}
\label{fig:alg-example1}
\vspace{-.1in}
\end{center}
\end{figure}
Consider the graph in Figure~\ref{fig:alg-example1} with its two limit cycle attractors, shown on the right.   To associate a sequence to the first attractor, notice that starting at a blue peak (neuron 1), next is always a green peak  (neuron 2), then a grey peak (neuron 3), followed by a low red peak (neuron 5) that is simultaneous with a yellow peak (neuron 4, not visible).  The corresponding sequence is thus 123(\underline{45}), where the (45) indicates that neurons 4 and 5 fire synchronously, and the underlining denotes low firing.  The second limit cycle in Figure~\ref{fig:alg-example1} has sequence 23\underline{51}4, with neurons 5 and 1 being low firing.

\paragraph{Matlab Exploration: Sequences of attractors\\}  
\hspace{-.2in} 
The Matlab package CTLN Basic, available at:
\begin{center}\url{https://github.com/nebneuron/CTLN-bookchapter}\end{center}
contains Matlab code to run simulations for CTLNs obtained from any directed graph.  Graphs can be coded by the user into the executable file, run\_CTLN\_model\_script.m, in the form of a binary adjacency matrix called sA (see README file for instructions).  The parameters $\varepsilon, \delta,$ and $\theta$ can also be adjusted in this file, with defaults matching those in this chapter. Initial conditions may be chosen at random, or specified by the user. This package was used to produce the firing rate plots seen in Figure~\ref{fig:alg-example1}, as well as those in earlier figures.

\begin{figure}[!ht]
\vspace{-.05in}
\begin{center}
\includegraphics[height=1.3in]{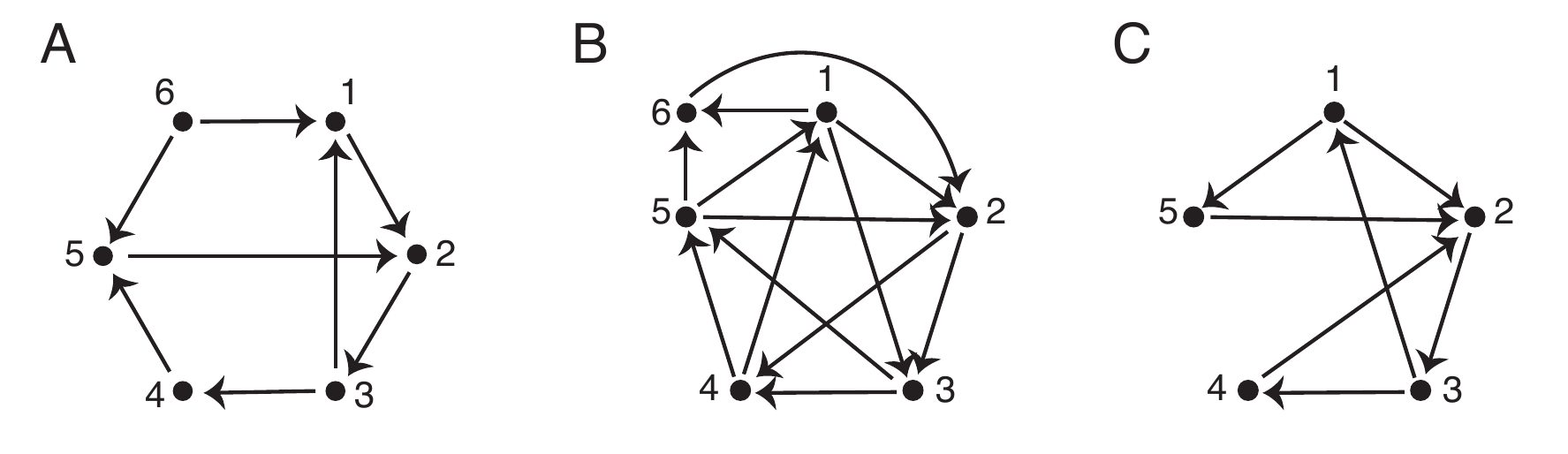}
\vspace{-.15in}
\caption{Graphs for exploratory Matlab activity. }
\label{fig:Matlab-examples}
\vspace{-.15in}
\end{center}
\end{figure}

Using CTLN Basic, code the graph in Figure~\ref{fig:Matlab-examples}A.  Try a variety of initial conditions to find the two limit cycle attractors of the corresponding CTLN.  Note that it may be helpful to try initial conditions where the non-zero entries correspond to a cycle in the graph.  Record the sequence of neural firing in each of the limit cycles.  Repeat this process for the two remaining graphs in Figure~\ref{fig:Matlab-examples}. \emph{Hint:} graph B has one attractor while graph C has two attractors.

%%%%%%%%%%%%%%%%%%%
\vspace{-.1in}
\subsection{Sequence prediction algorithm}
\vspace{-.05in}
Using the sequence prediction algorithm summarized in the table below, it is often possible to predict the sequences of attractors directly from the graph. This algorithm was first introduced in \cite{caitlyn-thesis}, but has been updated here.  To understand the algorithm, we first need some terminology.  

\vspace{-.05in}
\begin{definition} 
A node $k$ in $G$ is said to be \emph{freely removable} if removing $k$ from $G$

\vspace{-.05in}
\begin{enumerate}
\item[(a)] does not create a sink in $G\setminus\{k\}$ and

\vspace{-.05in}
\item[(b)] does not create a target-free 3-cycle in $G\setminus\{k\}$ if $k$ was otherwise a target of that 3-cycle.
\end{enumerate}

\vspace{-.05in}
\noindent We say a graph $G$ is \emph{irreducible} if it is cyclically symmetric or has no freely removable nodes.  If $G$ is cyclically symmetric, it is also called a \emph{core cycle}.  
\end{definition}

The sequence prediction algorithm consists of two phases; in the first phase, the graph $G$ is deconstructed by removing nodes according to the rules in the table until an irreducible subgraph is reached.  There may be choices as to which node to remove at a given step; each of these choices must be separately pursued to determine if different irreducible subgraphs result.  At the end of phase I, after pursuing all choices of nodes, a list of irreducible subgraphs is produced.  For each of these subgraphs that is a core cycle, the algorithm proceeds to phase II.  A sequence is then constructed for each core cycle as described in the previous table.  For each resulting sequence, we expect to see a corresponding attractor of the CTLN defined by $G$.  

\fbox{
\begin{minipage}{40em}
\medskip
{\bf Sequence prediction algorithm} (for an oriented graph $G$ with no sinks)\\

\vspace{-.1in}
\emph{Phase I: Deconstruct graph\\}

\vspace{-.15in}
If a node is removed at any step, return to Step 0 with the subgraph remaining after the node's removal.  If no node can be removed, then proceed to the next step.  Proceed to Phase II whenever the remaining subgraph is irreducible.  
\begin{enumerate}
\item[0.] Remove all sources (in-degree 0) from the graph.
\item Remove a node of in-degree 1 whose removal does not create a sink in the resulting subgraph.  
\item Remove a node of lowest in-degree that is \emph{freely removable}.  
\end{enumerate}
Output a list of core cycles of the graph.  If a final subgraph is irreducible, but not a core cycle, then declare algorithm failure for that subgraph.  \\

\emph{Phase II: Reconstruct sequence\\}

\vspace{-.15in}
Each core cycle yields a sequence, which is obtained as follows.
\begin{enumerate}
\item[0.]  List the core-cycle nodes in the order they cyclically appear. These are the high-firing nodes of the sequence. The inserted nodes (next step) will be low firing and thus underlined in the sequence.
\item For each node $i$ not in the core cycle, insert it into the sequence only if it receives at least one edge from the core cycle, as follows:
\begin{enumerate}
\item  Consider the induced subgraph $G|_\omega$ of core-cycle nodes $\omega$ that are inputs to node $i$. Insert $i$ in the sequence after each core-cycle node that is a sink in $G|_\omega$. 
\item If two nodes $i$ and $j$ are to be inserted in the same place, check how they interact.  If $i \to j$ in $G$, then insert $i$ before $j$, and vice versa if $j \to i$.  If there is no edge between $i$ and $j$, then the nodes will fire synchronously, denoted $(ij)$.  
\end{enumerate}
\end{enumerate}

\smallskip
\end{minipage}
}
\vspace{.2in}

\begin{example}
As an illustration of the sequence prediction algorithm, first consider the graph in Figure~\ref{fig:Matlab-examples}A.  The algorithm begins by removing the source node 6 (see Figure~\ref{fig:seq-algorithm}).  Then nodes 1, 3, 4, and 5 have in-degree 1; but neither 3 nor 5 can be removed because they would cause nodes 2 and 4, respectively, to become sinks.  When node 1 is removed, the remaining subgraph is a 4-cycle, which is cyclically symmetric, so $(2345)$ is irreducible and is a core cycle.  Alternatively, when node 4 is removed, node 5 becomes a source, and must be removed next.  Then the cycle $(123)$ remains, and this is a core cycle.  Thus, phase I outputs two core cycles $(2345)$ and $(123)$.

We now proceed to phase II.  Since node 6 was a source, it does not receive from any core cycle and is thus not inserted into any sequence. For the core cycle $(2345)$, the high-firing nodes are, in order, 2345 (or a cyclically equivalent ordering). Node 1 receives from the core-cycle node 3 only, and so it is inserted after 3 and underlined to indicate low firing. The predicted sequence is thus $23\underline{1}45$.  For the core cycle $(123)$, the high-firing part of the sequence is 123.  Node 4 receives from 3 only, and so is inserted after 3 in the sequence.  Node 5 does not receive from this core cycle, so it is \underline{not} inserted in this sequence. The second predicted sequence is thus $123\underline{4}$.  \\

\begin{figure}[!ht]
\vspace{-.05in}
\begin{center}
\includegraphics[height=2.3in]{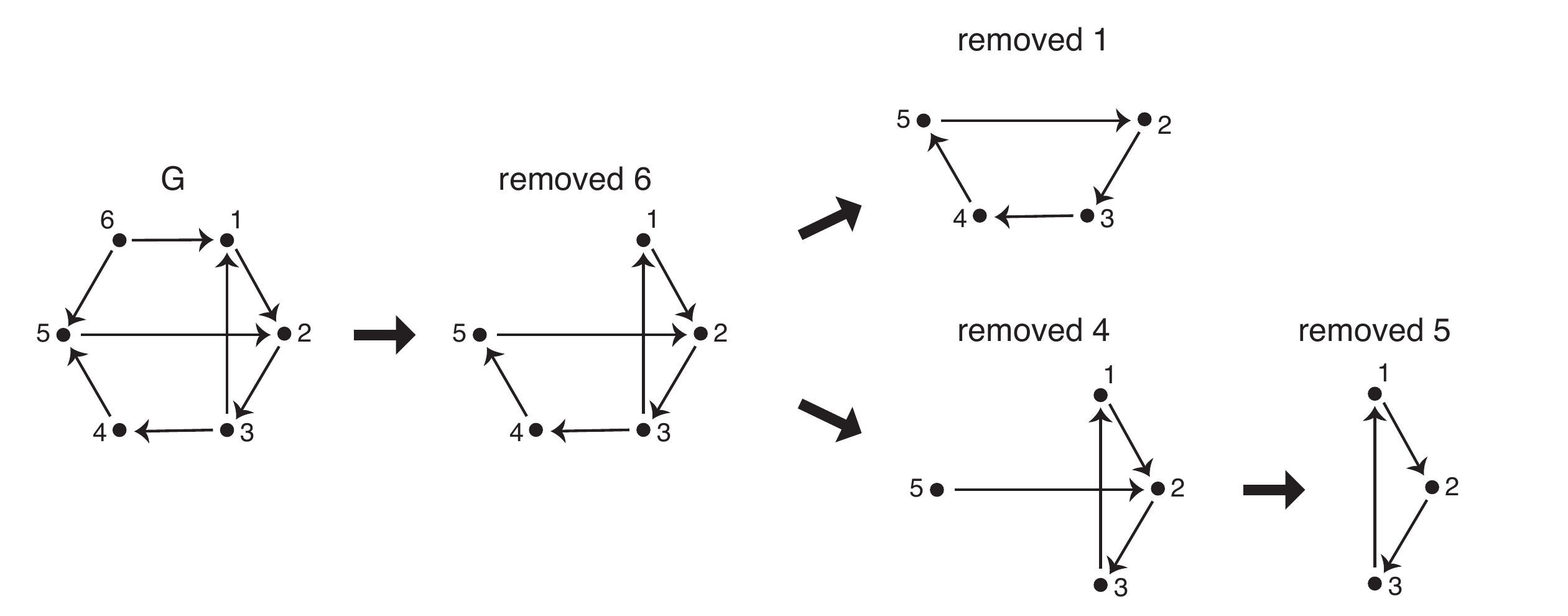}
\vspace{-.15in}
\caption{The graph in Figure~\ref{fig:Matlab-examples}A, deconstructed via phase I of the sequence prediction algorithm.}
\label{fig:seq-algorithm}
\vspace{-.15in}
\end{center}
\end{figure}

Next, consider the graph in Figure~\ref{fig:Matlab-examples}B.  There are no sources and no nodes of in-degree 1, so we proceed to Step 2.  Every node other than 2 has in-degree 2, and every node is a target of a 3-cycle (e.g.\ node 1 is a target of $(245)$).  Recall that a target of a 3-cycle is only freely removable if the 3-cycle has at least one other target. The only 3-cycle with two or more targets is $(135)$, whose targets are
nodes 2 and 6.  Since these nodes are not targets for any other 3-cycle, they are both freely removable.  We remove node 6, however, because it has lower in-degree. The remaining subgraph is cyclically symmetric, and thus is irreducible and a core cycle.  In phase II, we obtain the high-firing sequence 12345 from the core cycle. Node 6 receives from the core-cycle nodes $\omega = \{1,5\}$, but only $1$ is a sink of $G|_\omega$, so node 6 is inserted after $1$ only. This produces the sequence $1\underline{6}2345$.  \\

\vspace{-.1in}
Note that the sequences obtained above match those observed in the Matlab Exploration exercise.

\begin{exercise}
Perform the sequence prediction algorithm on graph C of Figure~\ref{fig:Matlab-examples} to obtain the two sequences found in the opening Matlab Exploration.
\end{exercise}

Finally, consider the graph in Figure~\ref{fig:alg-example1}.  There are no sources to remove, but nodes 3, 4, and 5 all have in-degree 1.  Removing node 3 would cause node 2 to become a sink, but both nodes 4 and 5 are valid candidates for removal.  If node 4 is removed first, then node 5 must be removed next, yielding the core cycle $(123)$.  If instead, node 5 is removed first, then either 1 or 4 can be removed next.  Removing node 4 produces the same $(123)$ core cycle already observed.  In contrast, removing node 1 produces a second core cycle $(234)$.  

In phase II, for the core cycle $(123)$, 123 is the sequence of high-firing neurons.  Both nodes 4 and 5 must be inserted into the sequence following node 3.  Since there is no edge between 4 and 5, these nodes will fire synchronously, producing the sequence 123(\underline{45}).  For the core cycle $(234)$, 234 is the sequence of high-firing neurons.  Nodes 1 and 5 must be inserted into the sequence after node 3.  Since $5 \to 1$ in $G$, node 5 is inserted first, yielding 23\underline{51}4.  Note that both these sequences have corresponding limit cycle attractors when $\varepsilon=0.35$ and $\delta=0.9$, as observed in Figure~\ref{fig:alg-example1}.  Interestingly, only the first attractor with sequence 123(\underline{45}) is observed for the standard parameters ($\varepsilon=0.25$ and $\delta=0.5$).
\end{example}

\begin{exercise}
Perform the sequence prediction algorithm on the graph in Figure~\ref{fig:intro-examples}D to obtain the two sequences shown there.  
\end{exercise}

The sequence prediction algorithm has been tested on all 160 permutation-inequivalent oriented graphs with no sinks on $n\leq 5$ neurons \cite{caitlyn-thesis}.  With the standard parameters, the algorithm correctly predicted the sequences corresponding to all the attractors of the CTLNs for 152 of the 160 graphs.  There were four graphs for which multiple sequences were predicted, but each had one sequence that was not observed with the standard parameters; for example, the second sequence in Figure~\ref{fig:alg-example1}.  
Nevertheless, for each of these graphs there was an alternative set of legal parameters where all predicted sequences were observed.  Thus, taken across the full legal parameter range, the algorithm was successful for 156 of the 160 graphs \cite{caitlyn-thesis}.  

\begin{figure}[!ht]
\begin{center}
\includegraphics[height=1.65in]{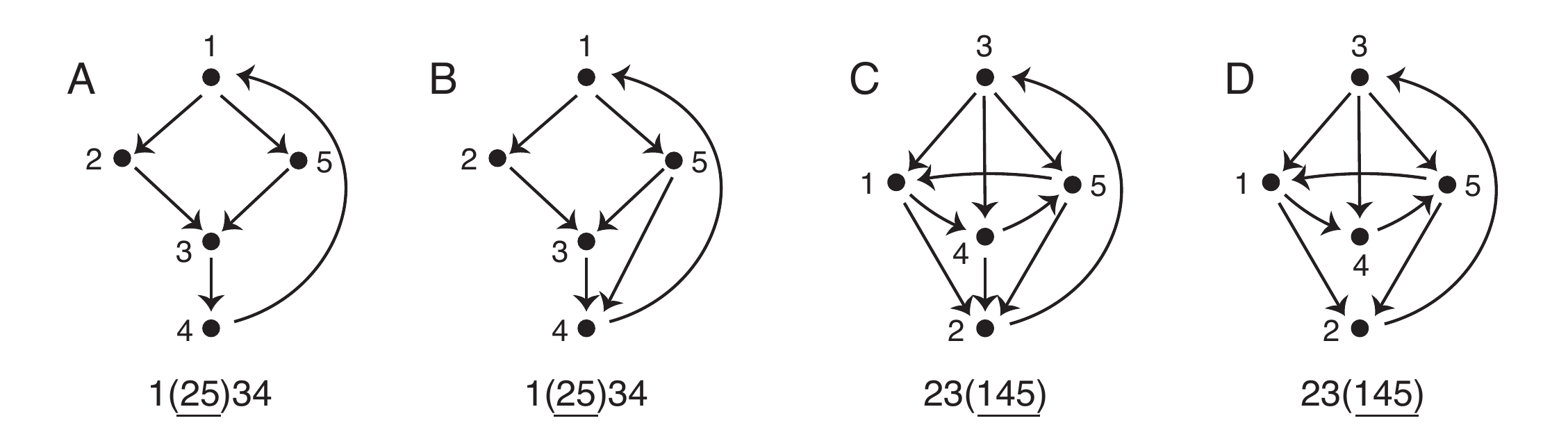}
\vspace{-.2in}
\caption{Graphs for which the sequence prediction algorithm fails across all legal parameter values. Below each graph is the sequence of the single attractor that is observable across all legal parameters.}
\label{fig:alg-failures}
\vspace{-.25in}
\end{center}
\end{figure}

There were four graphs, however, for which the algorithm consistently failed to predict the sequences; these are shown in Figure~\ref{fig:alg-failures} along with the actual observed sequences.  The attractors for these graphs all have synchrony that is not predicted by the algorithm, and which would require merging multiple core cycles.  This unexpected synchrony is the result of \emph{graph automorphisms} in graphs A and C, while graphs B and D are one edge away from having a graph automorphism.  The problem with the sequence prediction, therefore, appears to arise from symmetry that is not taken into account by the algorithm. In the next section, we explore graph automorphisms in more detail and examine their impact on the set of attractors of a network.

%%%%%%%%%%%%%%%%%%%
\subsection{Symmetry of graphs acting on the space of attractors}
Informally, a graph automorphism is a bijective map from a graph to itself that reflects its symmetry.  More precisely, for a graph $G$ with vertex set $V(G)$, a bijection $\alpha: V(G) \to V(G)$ is a \emph{graph automorphism} if it preserves the edges of $G$, i.e.\ $i \to j$ in $G$ if and only if $\alpha(i) \to \alpha(j)$.  

\begin{example}
The butterfly graph from Figure~\ref{fig:butterfly-survival}A has a graph automorphism $\alpha$ that interchanges nodes 1 and 4 and fixes nodes 2 and 3.  This is an automorphism since the map sends the edges $1 \to 2$ and $4 \to 2$ to each other and similarly sends the edges $ 3 \to 1$ and $3 \to 4$ to each other, while fixing the edge $2 \to 3$.  
\end{example}

\begin{example}
For graph A in Figure~\ref{fig:alg-failures}, the map $\alpha$ that sends node 2 to 5 and vice versa, while fixing all other nodes, is a graph automorphism.  This same map is \underline{not} an automorphism of graph B since  it sends the edge $5 \to 4$ to the edge $2 \to 4$, which is not present in the original graph.  

For graph C in Figure~\ref{fig:alg-failures}, the map $\alpha$ that sends node 1 to 4, 4 to 5, and 5 to 1, while fixing nodes 2 and 3, is a graph automorphism. (Verify this!) 
Note that $\alpha$ is not an automorphism of graph D because the edge $1 \to 2$ is mapped to $4 \to 2$, which is not an edge of the original graph.  
\end{example}

\begin{exercise}\label{ex:automorphism}
Return to the graphs in Figure~\ref{fig:activity-stable-fp}.  Identify which graphs have nontrivial graph automorphisms and find the automorphisms of these graphs.
\end{exercise}

How does a graph automorphism affect the corresponding CTLN? Since the map permutes the neurons of a network in a way that preserves the graph, the resulting CTLN must have the same set of attractors. In other words, the automorphism induces a bijection on the set of attractors. A single attractor may be fixed, or sent to another attractor that differs from it only by permuting neuron labels.
For example, the attractor of a 3-cycle $(123)$ has neurons 1, 2, and 3 periodically firing in sequence, and the graph automorphism that sends node 1 to 2, 2 to 3, and 3 to 1 fixes the attractor.  Another way an attractor can be fixed is if the exchanged nodes in the automorphism fire synchronously. For example, the attractor produced by the graph in Figure~\ref{fig:alg-failures}A has neurons 2 and 5 firing synchronously, and thus the 
 automorphism that exchanges nodes 2 and 5 maps this attractor to itself.  Similarly, the automorphism of the graph in Figure~\ref{fig:alg-failures}C fixes the attractor, since the exchanged nodes 1,4, and 5 fire synchronously.
 
 \begin{exercise}[Extended research project]
Investigate graphs with automorphisms and find the set of attractors of their CTLNs using the Matlab package provided. It is easiest to find all attractors by choosing initial conditions that are small perturbations around the unstable fixed points of the network.  Compare the sequences of attractors observed to the outputs of the sequence prediction algorithm.  Investigate possible modifications to the sequence prediction algorithm to improve its predictive power for graphs with various types of automorphisms.  Test your improved algorithm on the graphs in Figure~\ref{fig:alg-failures}.
\end{exercise}

When an automorphism does \underline{not} fix an attractor, it must send it to another one of the same type. For example, the butterfly graph in Figure~\ref{fig:butterfly-survival}A has an automorphism that exchanges nodes 1 and 4. The two attractors of this network are identical up to permutation, and have sequences $123\underline{4}$ and $423\underline{1}$. It is easy to see that the automorphism sends these attractors to each other.  Note that if we had only discovered one of them, the automorphism would tell us that the other must exist.
In this way, the presence of a graph automorphism can aid in predicting new attractors in a network once a (non-fixed) one has been observed.  

\begin{example}
Consider the graph in Figure~\ref{fig:phone-number}, with $m$ neurons (nodes) in each layer, where the layers wrap around in a cylindrical fashion. This graph has two types of automorphisms. The first type consists of all permutations of the nodes that keep each node inside its original layer. If we consider the limit cycle displayed on the right, we see that each such automorphism produces another limit cycle with a different set of 5 neurons firing in sequence. The existence of one such limit cycle, together with the $m^5$ automorphisms, thus predicts that this network has at least $m^5$ sequential attractors.
Note that this network architecture, with 7 layers, could serve as a mechanism for storing phone numbers in working memory ($m = 10$ for digits $0-9$). The phone number is stored as a sequence that is repeated indefinitely, with different initial conditions producing different phone number sequences. 
(Can you find another type of automorphism for this network? What does it do to the attractors?)

\begin{figure}[!ht]
\begin{center}
\includegraphics[width=.9\textwidth]{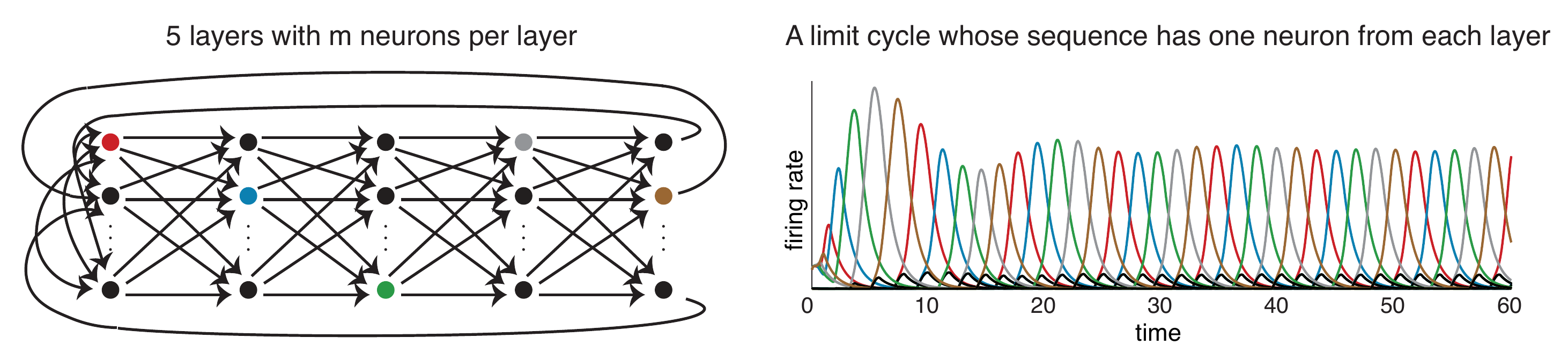}
\vspace{-.1in}
\caption{(Left) A cyclically structured graph with $m$ neurons per layer, and all $m^2$ feedforward connections from one layer to the next. (Right) A limit cycle for the corresponding CTLN (with parameters 
$\varepsilon=0.75$, $\delta=4$) in which 5 neurons fire in a repeating sequence, with one neuron from each layer.}
\label{fig:phone-number}
\vspace{-.3in}
\end{center}
\end{figure}
\end{example}

\begin{figure}[!ht]
\begin{center}
\includegraphics[width=.985\textwidth]{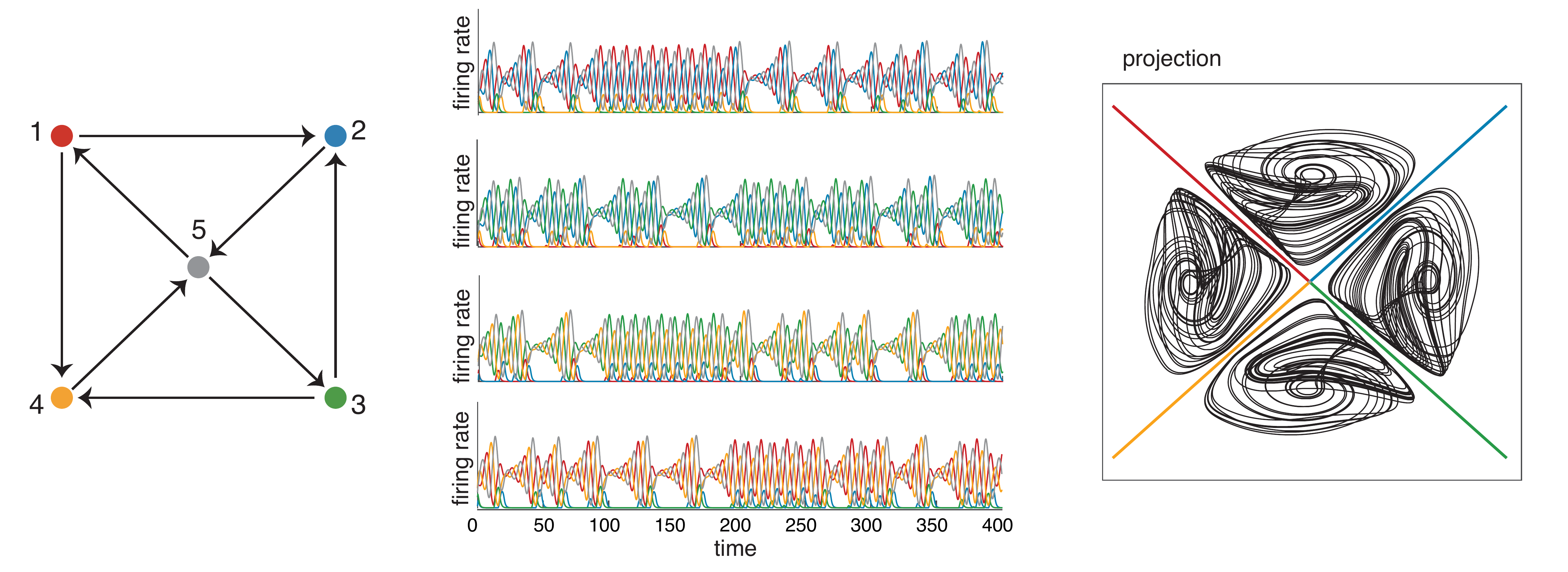}
\vspace{-.15in}
\caption{Graph of CTLN with four chaotic attractors -- one attractor for each 3-cycle.  Note that this graph differs from the graph in Figure~\ref{fig:intro-examples}B only by flipping the $3 \to 4$ edge.}
\label{fig:baby-chaos}
\vspace{-.15in}
\end{center}
\end{figure}

Graph automorphisms may also play a role in producing more exotic attractors.   
For example, in Figure~\ref{fig:baby-chaos} we see chaotic attractors in a network of only 5 nodes, with four perfectly symmetric overlapping 3-cycles.  There is an attractor for each 3-cycle, and these attractors are chaotic for the standard parameters.  Interestingly, for different (legal) parameter values the chaotic attractors may become limit cycles. The graph automorphisms ($1 \leftrightarrow 3$ and $2 \leftrightarrow 4$) permute these attractors without fixing any.

\begin{figure}[!ht]
\vspace{-.1in}
\hspace{-.15in}\includegraphics[width=1.05\textwidth]{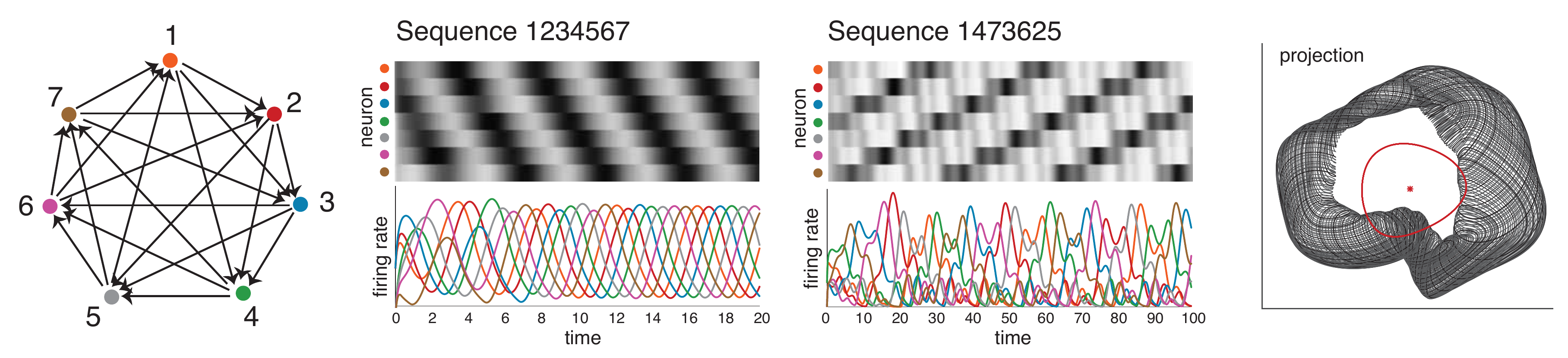}
\vspace{-.4in}
\begin{center}
\caption{(Left) Graph of a cyclically symmetric CTLN with two attractors: one limit cycle and one quasiperiodic. (Middle) The limit cycle has the expected sequence 1234567, while the quasiperiodic attractor has a sequence corresponding to another cycle in the graph. (Right) A $2$-dimensional projection of the two attractors, and the unique (unstable) fixed point. The limit cycle and fixed point are shown in red, while the quasiperiodic attractor is the torus-like trajectory shown in black.}
\label{fig:7-star}
\vspace{-.35in}
\end{center}
\end{figure}

We also see quasiperiodic attractors in small networks with symmetry.  A \emph{quasiperiodic} attractor is one that is nearly periodic, but whose trajectory does not perfectly repeat.  The attractor thus has the shape of a torus, rather than a circle as in the case of a limit cycle.  Figure~\ref{fig:7-star} displays a CTLN from a cyclically symmetric graph on $n=7$ nodes.  For standard parameters, this network has two attractors: one limit cycle and one quasiperiodic. Interestingly, the quasiperiodic attractor only emerges for a portion of the legal parameter range, while the limit cycle is present for the full range.  The emergence of the quasiperiodic attractor is particularly surprising because it does not have a corresponding unstable fixed point.  In fact, this network has exactly one unstable fixed point, and initial conditions near this fixed point always yield the limit cycle. The quasiperiodic attractor can be obtained from a much smaller set of initial conditions, including $[0.1, 0, 0, 0.1, 0, 0, 0]$.

\begin{exercise}[Mini project, for those familiar with bifurcation theory] Conduct a bifurcation analysis for the graph in Figure~\ref{fig:alg-example1}.  Fix $\varepsilon>0$ and choose a set of $\delta$s ranging from $-\varepsilon$ to 2.  For each choice of parameters, find the collection of attractors of the CTLN by sampling a variety of initial conditions, including perturbations of all unstable fixed points.  Identify the parameter values where bifurcations occur by finding when the set of dynamic attractors qualitatively changes.  
\end{exercise}

\section*{Acknowledgments}
\vspace{-.1in}
KM and CC were supported by NIH R01 EB022862.  CC was also supported by NSF DMS-1516881.  We thank David Falk, an undergraduate student, for his help in developing various exercises and examples.

%%%%%%%%%%%%%%%%%%%
\bibliographystyle{unsrt}
\bibliography{CTLN-refs}

\begin{thebibliography}{10}

\bibitem{AppendixE}
H.S. Seung and R.~Yuste.
\newblock {\em Principles of Neural Science}, chapter Appendix {E}: Neural
  networks, pages 1581--1600.
\newblock McGraw-Hill Education/Medical, 5th edition, 2012.

\bibitem{CTLN-paper}
K.~Morrison, J.~Geneson, C.~Langdon, A.~Degeratu, V.~Itskov, and C.~Curto.
\newblock Emergent dynamics from network connectivity: a minimal model.
\newblock \emph{In preparation.} Earlier version available online at
  \verb!https://arxiv.org/abs/1605.04463!

\bibitem{Hopfield1}
J.J. Hopfield.
\newblock Neural networks and physical systems with emergent collective
  computational abilities.
\newblock {\em Proc. Natl. Acad. Sci.}, 79(8):2554--2558, 1982.

\bibitem{Amit-ANNs}
Daniel~J. Amit.
\newblock {\em Modeling brain function: {T}he world of attractor neural
  networks}.
\newblock Cambridge University Press, 1989.

\bibitem{Marder-CPG}
E.~Marder and D.~Bucher.
\newblock Central pattern generators and the control of rhythmic movements.
\newblock {\em Curr. Bio.}, 11(23):R986--996, 2001.

\bibitem{CPG-models}
J.J. Collins and I.N. Stewart.
\newblock Coupled nonlinear oscillators and the symmetries of animal gates.
\newblock {\em J. Nonlinear Sci.}, 3:349--392, 1993.

\bibitem{ErmentroutTerman}
G.B. Ermentrout and D.H. Terman.
\newblock {\em Mathematical foundations of neuroscience}.
\newblock Springer-Verlag New York, 2010.

\bibitem{Stark-PNAS}
E.~Stark, L.~Roux, R.~Eichler, and G.~Buzs\'{a}ki.
\newblock Local generation of multineuronal spike sequences in the hippocampal
  {CA}1 region.
\newblock {\em Proc. Natl. Acad. Sci.}, 112(33):10521--10526, 2015.

\bibitem{Eva-Science}
E.~Pastalkova, V.~Itskov, A.~Amarasingham, and G.~Buzs\'{a}ki.
\newblock Internally generated cell assembly sequences in the rat hippocampus.
\newblock {\em Science}, 321(5894):1322--1327, 2008.

\bibitem{Luczak-PNAS}
A.~Luczak, P.~Barth\'{o}, S.L. Marguet, G.~Buzs\'{a}ki, and K.D. Harris.
\newblock Sequential structure of neocortical spontaneous activity {\em in
  vivo}.
\newblock {\em Proc. Natl. Acad. Sci.}, 104(1):347--352, 2007.

\bibitem{Buzsaki}
G.~Buzs\'{a}ki.
\newblock {\em Rhythms of the brain}.
\newblock Oxford University Press, 2011.

\bibitem{HahnSeungSlotine}
R.~H. Hahnloser, H.S. Seung, and J.J. Slotine.
\newblock Permitted and forbidden sets in symmetric threshold-linear networks.
\newblock {\em Neural Comput.}, 15(3):621--638, 2003.

\bibitem{pattern-completion}
C.~Curto and K.~Morrison.
\newblock Pattern completion in symmetric threshold-linear networks.
\newblock {\em Neural Computation}, 28:2825--2852, 2016.

\bibitem{fp-paper}
C.~Curto, J.~Geneson, and K.~Morrison.
\newblock Fixed points of competitive threshold-linear networks.
\newblock Available online at \verb!https://arxiv.org/abs/1804.00794/!

\bibitem{moon-moser}
J.~W. Moon and L.~Moser.
\newblock On cliques in graphs.
\newblock {\em Israel J. Math.}, 3:23--28, 1965.

\bibitem{poincare-hopf}
J.~W. Milnor.
\newblock {\em Topology from the differentiable viewpoint}.
\newblock Princeton Landmarks in Mathematics. Princeton University Press,
  Princeton, NJ, 1997.

\bibitem{caitlyn-thesis}
C.~Parmelee.
\newblock {\em Applications of dicrete mathematics for understanding dynamics
  of synapses and networks in neuroscience}.
\newblock PhD thesis, University of Nebraska-Lincoln, 2016.

\end{thebibliography}
%
%%%%%%%%%%%%%%%%%%%%
\section*{Appendix: Review of linear systems of ODEs}
In this appendix, we briefly review linear systems of differential equations with constant coefficients; for further details, see any textbook on ordinary differential equations.  

Consider a linear system with constant real-valued coefficients of the form

\vspace{-.15in}
{\small \begin{eqnarray*}
\frac{dx_1}{dt} &=& a_{11}x_1 + a_{12}x_2 + \ldots + a_{1n}x_n + b_1\\
\frac{dx_2}{dt} &=& a_{21}x_1 + a_{22}x_2 + \ldots + a_{2n}x_n + b_2\\
& \vdots & \\
\frac{dx_n}{dt} &=& a_{n1}x_1 + a_{n2}x_2 + \ldots + a_{nn}x_n + b_n\\
\end{eqnarray*}
}

\vspace{-.4in}
\noindent which can be written compactly as

\vspace{-.3in}
\begin{eqnarray}\label{eq:system}
\frac{d\vx}{dt} = A\vx + \vb
\end{eqnarray}

\vspace{-.1in}

\noindent where $\vx$ and $\vb$ are column vectors and $A$ is the matrix of coefficients.
A \emph{fixed point} of such a system is a point $\vx^*$ where each derivative $dx_i/dt$ is zero, i.e.\ $\frac{d\vx}{dt}|_{\vx = \vx^*}=0$.  Assuming the matrix $A$ is invertible, the system \eqref{eq:system} has a unique fixed point $\vx^* = -A^{-1}\vb$.  The behavior of the system around this fixed point is dictated by the eigenvalues $\{\lambda_i\}$ and corresponding eigenvectors $\{\vv_i\}$ of the matrix $A$.  \\

\begin{figure}[!ht]
\begin{center}
\includegraphics[width=1\textwidth]{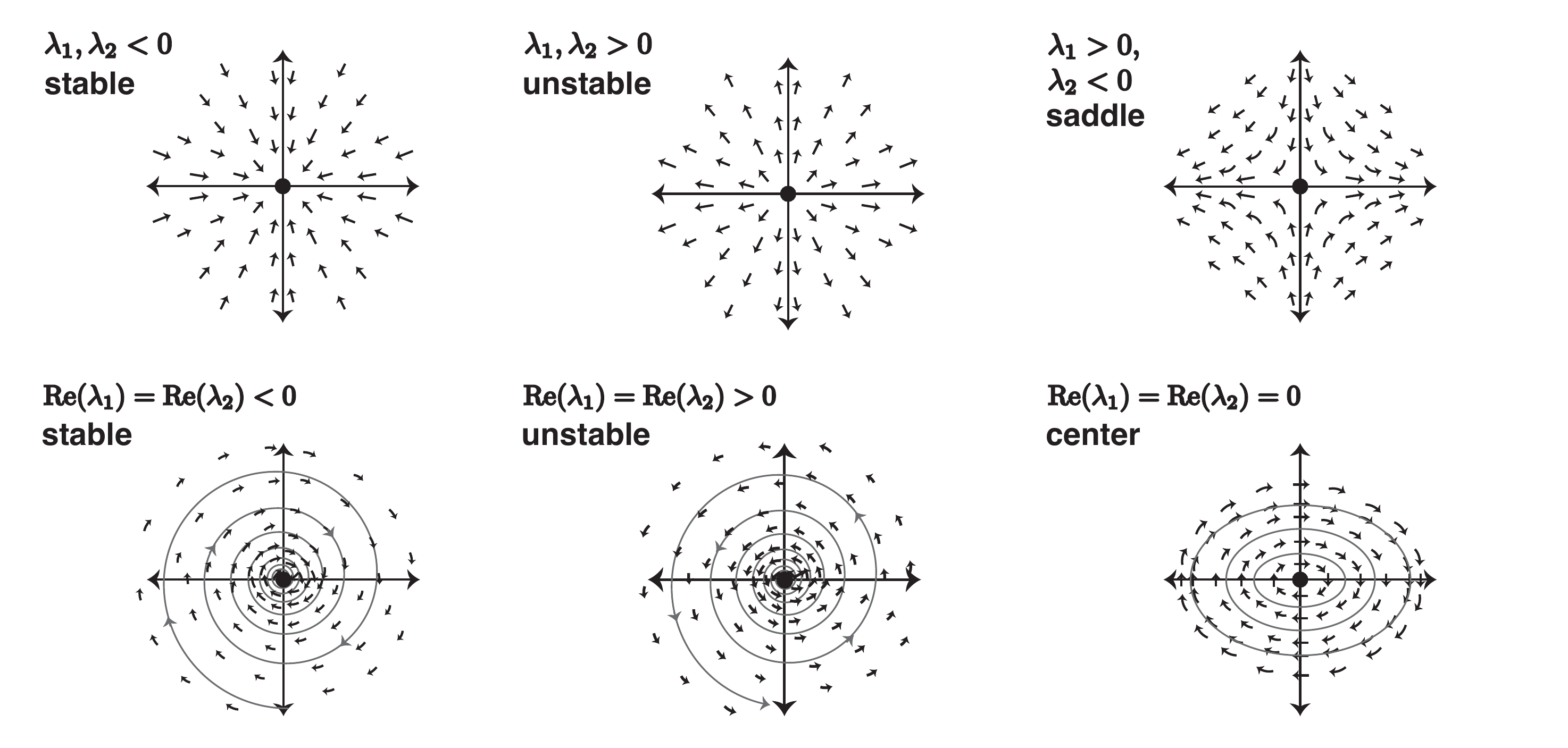}
\vspace{-.2in}
\caption{Sample vector fields for each of the six types of fixed points that can arise in a two-dimensional linear system.}
\label{fig:linear-systems}
\end{center}
\vspace{-.15in}
\end{figure}

For simplicity, consider the system \eqref{eq:system} with just two variables $\vx= {\footnotesize \begin{bmatrix} x_1\\  x_2 \end{bmatrix}}$ and $2 \times 2$ matrix $A$, with eigenvalues $\lambda_1$ and $\lambda_2$ corresponding to eigenvectors $\vv_1$ and $\vv_2$, respectively.  When the eigenvalues are distinct, the solutions to the system have the form 
$$\vx(t) = C_1\vv_1e^{\lambda_1 t} + C_2\vv_2e^{\lambda_2 t} -A^{-1}\vb,$$  
where $C_1, C_2$ are constants determined by the initial conditions.  Observe that if $\lambda_1$ and $\lambda_2$ are both real and negative, then the first two terms decay towards 0 as $t \to \infty$ and all trajectories converge to the fixed point $\vx^*=-A^{-1}\vb$. In this case, the fixed point is \emph{stable} and is an \emph{attractor} of the network.  If the eigenvalues are both real and positive, then solutions will tend toward infinity as $t \to \infty$, and the fixed point $\vx^*$ is called \emph{unstable}. In contrast, if one eigenvalue is positive while the other is negative, then the fixed point is a \emph{saddle}. 
The top row of Figure~\ref{fig:linear-systems} shows vector fields centered at the fixed point with axes corresponding to the eigenvectors for each of these three cases.  For ease of drawing, we have assumed the eigenvalues are distinct and the eigenvectors are orthogonal.

When the eigenvalues are complex, $e^{\lambda_i t}$ can be rewritten as $e^{\operatorname{Re}(\lambda_i)t}[\cos(\operatorname{Im}(\lambda_i)t) + i \sin(\operatorname{Im}(\lambda_i)t)]$ using Euler's formula. This produces spiral-like behavior that converges toward the fixed point when $\operatorname{Re}(\lambda_i)<0$ and diverges when $\operatorname{Re}(\lambda_i)>0$. (Note that $\operatorname{Re}(\lambda_1) =  \operatorname{Re}(\lambda_2)$, as the eigenvalues must be complex conjugate pairs.)  In the former case, the fixed point is again called \emph{stable}, while in the latter case it is \emph{unstable}.  Finally, when $\operatorname{Re}(\lambda_i)=0$, the system produces perfectly periodic orbits around the fixed point, but these are unstable, as small changes in initial conditions result in different periodic trajectories.  Vector fields corresponding to these three cases are shown in the bottom panel of Figure~\ref{fig:linear-systems}.  

In higher-dimensional linear systems, these same behaviors occur around fixed points, and the fixed point's type is again dictated by the eigenvalues of the matrix $A$.  A fixed point of the system \eqref{eq:system} is \emph{stable} when all eigenvalues have negative real part, and is {\it unstable} if at least one eigenvalue has positive real part.

\end{document}